\documentclass[fleqn,usenatbib]{mnras}

\usepackage{amssymb}	
 
\usepackage{newtxtext,newtxmath}

\usepackage[T1]{fontenc}

\DeclareRobustCommand{\VAN}[3]{#2}
\let\VANthebibliography\thebibliography
\def\thebibliography{\DeclareRobustCommand{\VAN}[3]{##3}\VANthebibliography}

\usepackage{graphicx}	
\usepackage{amsmath}	
\usepackage{siunitx}
\usepackage{float}
\usepackage{makecell}
\usepackage{amsfonts,mathbbol}
\usepackage{physics}
\usepackage{threeparttable}

\newcommand{\chandra}{\textit{Chandra}\ }
\DeclareSIUnit\erg{erg}
\DeclareSIUnit\parsec{pc}
\DeclareSIUnit\jansky{Jy}
\mathchardef\mhyphen="2D

\title[First limits on the BHs in NGC~3201]{The MAVERIC Survey: The first radio and X-ray limits on the detached black holes in NGC~3201}

\author[Alessandro Paduano et al.]
{Alessandro Paduano$^{1}$\thanks{E-mail: a.paduano@postgrad.curtin.edu.au},
Arash Bahramian$^{1}$,
James C. A. Miller-Jones$^{1}$,
Adela Kawka$^{1}$,
\newauthor Fabian G\"ottgens$^{2}$,
Jay Strader$^{3}$,
Laura Chomiuk$^{3}$,
Sebastian Kamann$^{4}$,
Stefan Dreizler$^{2}$,
\newauthor Craig O. Heinke$^{5}$,
Tim-Oliver Husser$^{2}$,
Thomas J. Maccarone$^{6}$,
Evangelia Tremou$^{7}$,
\newauthor and Yue Zhao$^{5}$
\\
$^{1}$International Centre for Radio Astronomy Research - Curtin University, GPO Box U1987, Perth, WA 6845, Australia\\
$^{2}$Institut f\"ur Astrophysik, Georg-August-Universit\"at G\"ottingen, Friedrich-Hund-Platz 1, 37077 G\"ottingen, Germany\\
$^{3}$Center for Data Intensive and Time Domain Astronomy, Department of Physics and Astronomy, Michigan State University, East Lansing, MI 48824, USA\\
$^{4}$Astrophysics Research Institute, Liverpool John Moores University, 146 Brownlow Hill, Liverpool L3 5RF, UK\\
$^{5}$Department of Physics, University of Alberta, CCIS 4-181, Edmonton, AB T6G 2E1, Canada\\
$^{6}$Department of Physics \& Astronomy, Texas Tech University, Box 41051, Lubbock, TX 79409-1051, USA\\
$^{7}$LESIA, Observatoire de Paris, CNRS, PSL, SU/UPD, Meudon, France
}

\date{Accepted XXX. Received YYY; in original form ZZZ}

\pubyear{2021}

\begin{document}
\label{firstpage}
\pagerange{\pageref{firstpage}--\pageref{lastpage}}
\maketitle

\begin{abstract}
    The Galactic globular cluster NGC~3201 is the first Galactic globular cluster observed to host dynamically-confirmed stellar-mass black holes, containing two confirmed and one candidate black hole. This result indicates that globular clusters can retain black holes, which has important implications for globular cluster evolution. NGC~3201 has been observed as part of the MAVERIC survey of Galactic globular clusters. We use these data to confirm that there is no radio or X-ray detection of the three black holes, and present the first radio and X-ray limits on these sources. These limits indicate that any accretion present is at an extremely low rate and may be extremely inefficient. In particular, for the system ACS ID \#21859, by assuming the system is tidally locked and any accretion is through the capture of the companion's winds, we constrain the radiative efficiency of any accretion to $\lesssim\num{1.5e-5}$. We also combine the radio and X-ray source catalogues from the MAVERIC survey with the existing MUSE spectroscopic surveys and the HUGS catalogue of NGC~3201 to provide a catalogue of 42 multiwavelength sources in this cluster. We identify a new red straggler source with X-ray emission, and investigate the multiwavelength properties of the sub-subgiant population in the cluster.
\end{abstract}




\section{Introduction}
\subsection{X-ray binaries in globular clusters} \label{sec:intro_XRBsinGCs}
    X-ray binaries (XRBs) are binary systems that contain a compact object, either a black hole (BH) or neutron star (NS) accreting from a companion star. Low mass X-ray binaries (LMXBs) are XRBs where the mass of the companion star is $<1M_{\odot}$. Approximately one third of all known XRBs are believed to contain BHs as the compact object \citep{Liu2007,Corral-Santana2016,Tetarenko2016}. 
    
    Globular clusters (GCs) are large, gravitationally bound clusters of stars orbiting the Galactic centre. The Milky Way contains 156 known GCs \citep{Harris1996,Baumgardt2019} with stellar densities in these clusters reaching $10^6$ stars $\textrm{pc}^{-3}$. It has been shown observationally that GCs contain an overabundance of XRBs of two orders of magnitude per unit mass compared to the Galactic field \citep{Clark1975}. This overabundance is due to the formation channels of XRBs in GCs. In a cluster, XRBs will form dynamically, through channels including collisions between giant stars and compact objects \citep{Sutantyo1975}, tidal captures by compact objects \citep{Fabian1975}, and exchange interactions into primordial binaries \citep{Hills1976}. This is in contrast to the Galactic field, where binary evolution is the dominant mechanism for XRB formation. 
    
    GCs include many populations of X-ray emitting binaries. NS-LMXBs account for most of the bright systems \citep[$L_\textrm{X}>\SI{e34}{\erg\per\second}$, ][]{Bahramian2014}, and some sources in quiescence. Millisecond pulsars (MSPs), recycled pulsars that are spun up through accretion from a companion to millisecond spin periods \citep{Lorimer2008}, are abundant in GCs, with more than 150 sources detected\footnote{\url{http://www3.mpifr-bonn.mpg.de/staff/pfreire/GCpsr.html}}. Related to MSPs are their suspected progenitor systems, transitional millisecond pulsars (tMSPs). These systems are LMXBs that switch between a rotation-powered radio pulsar and a sub-luminous LMXB state \citep{Archibald2009,Papitto2013,Bassa2014}. There are currently three confirmed tMSPs and a handful of other candidates, of which one confirmed tMSP resides within the GC M28 \citep{Papitto2013}. Two candidate tMSPs also reside within the clusters Terzan 5 and NGC 6652 respectively \citep{Bahramian2018a,Paduano2021}. Cataclysmic variables (CVs), white dwarfs accreting from low mass stars, make up a large fraction of X-ray emitting binaries in globular clusters \citep{Grindlay2001,Pooley2002}. There is also a group of chromospherically active binaries (ABs) consisting of two tidally locked stars in a close binary that make up the majority of X-ray sources with luminosities $\lesssim\SI{e31}{\erg\per\second}$ \citep{Gudel2002}.
    
    Another group of rare binaries found in GCs are those with stripped companion stars. These systems are called sub-subgiants (SSGs) if these stars are fainter than subgiants but redder than a main sequence star when placed on a colour-magnitude diagram, and called red stragglers if they are brighter than subgiants but redder than normal giants \citep{Geller2017}. These stars have been observed to be X-ray sources \citep{vandenBerg1999}. A possible formation channel for these systems is through grazing tidal captures, resulting in a stripped, underluminous subgiant star \citep{Ivanova2017,Shishkovsky2018}. A total of 65 SSGs and red stragglers in 16 GCs have been identified \citep{Geller2017}, with 23 of these sources being faint ($L_\textrm{X}\sim10^{30-31}\si{erg\per\second}$) X-ray sources. Binary evolution is crucial in SSG formation, with this formation channel being the most prevalent, especially in GCs \citep{Geller2017,Geller2017a}. 
    
    Due to their age, GCs are expected to host a population of BHs. However, such a population has remained undetected for decades, which agreed with early theories about BH interactions within clusters. The BHs born from the deaths of the most massive stars in the cluster will sink to the centre of the cluster through mass segregation, with the Spitzer mass instability \citep{Spitzer1969} causing the BHs to form a small, dynamically decoupled sub-cluster. The mutual gravitational interactions within this sub-cluster would cause the BH population to be ejected from the cluster. Early analytical estimates and simulations of globular clusters indicated that a very small or non-existent BH population would remain \citep[e.g., ][]{Kulkarni1993,Sigurdsson1993,Heggie2008}.
    
    Observational studies over the past decade and a half began to cast doubt over the once held hypothesis that all BHs were ejected from the cluster. Several extra-galactic \citep[e.g., ][]{Maccarone2007,Shih2010,Brassington2010,Irwin2010,Barnard2011,Maccarone2011} and Galactic \citep[e.g., ][]{Strader2012a,Chomiuk2013,Miller-Jones2015,Shishkovsky2018,Zhao2020} BH XRB candidates in globular clusters have been identified, suggesting that a population of BHs can exist within globular clusters without being ejected. A simultaneous advance in numerical simulations indicated that the Spitzer mass instability is not valid and that only the most massive BHs sink to the centre of the cluster, leaving the remaining BHs well mixed within the cluster \citep{Mackey2008,Moody2009,Breen2013,Morscher2013,Morscher2015}, thus providing a theoretical basis for the presence of BHs within globular clusters. In 2018, decades of debate was resolved when a detached BH was dynamically detected within the Galactic GC NGC 3201 \citep{Giesers2018}, with another BH (and one candidate) detected the following year \citep{Giesers2019}, proving that a population of BHs can remain within GCs to the present day.

\subsection{NGC 3201}
    NGC 3201 is a GC with a mass of $\num{1.41e5} M_{\odot}$ located at a distance of $(4.74\pm0.04)$ \si{kpc} \citep{Baumgardt2019}. It is an extended cluster, with a core radius of 1.3 arcmin (1.22 pc) and a half-light radius of 3.1 arcmin (3.80 pc) \citep{Baumgardt2019}, and has a high binary fraction ($0.128\pm0.008$) within the core \citep{Milone2012}.

    NGC 3201 has recently become an important cluster for study of binary and BH dynamics in GCs, with the discovery of two detached BHs (and one candidate) in the cluster through observations using the Multi Unit Spectroscopic Explorer \citep[MUSE; ][]{Bacon2010}. A large survey of several Galactic GCs is being conducted with MUSE enabling the spectra of several thousand stars per cluster to be obtained \citep{Husser2016,Kamann2018}. Through radial velocity variations of sources within NGC 3201, \citet{Giesers2018,Giesers2019} detected two BHs of mass $4.53\pm0.21 M_{\odot}$ and $7.68\pm0.5 M_{\odot}$, and one candidate BH of mass $4.4\pm2.8 M_{\odot}$. All three sources are in detached binary systems, meaning there is very little (if any) mass transfer onto the compact object. Further analyses of the MUSE data have investigated the binary population within the cluster \citep{Giesers2019}, and have categorised various emission line sources \citep{Gottgens2019}.

\subsection{The MAVERIC Survey}
    The detection of the first BH candidates within the Galactic GC M22 by \citet{Strader2012a} spurred the creation of the Milky-way ATCA and VLA Exploration of Radio-sources in Clusters survey \citep[MAVERIC; ][]{Shishkovsky2020,Bahramian2020}. The MAVERIC sample consists of 50 GCs, primarily including all clusters that have a distance $<\SI{9}{kpc}$ and a mass $>10^5M_{\odot}$. Both the Karl G. Jansky Very Large Array (VLA) and the Australia Telescope Compact Array (ATCA) were used to observe this sample to produce the deepest radio survey of GCs. The VLA observations were conducted at 5.0 and 7.2 GHz, and the ATCA observations were conducted at 5.5 and 9.0 GHz. Further details about the observations of the cluster sample and the reduction of the radio data are presented in \citet{Tremou2018} and \citet{Shishkovsky2020}. In addition to the identification of several BH candidates in Galactic GCs (see~\ref{sec:intro_XRBsinGCs}), the MAVERIC survey has also identified tMSP candidates \citep{Bahramian2018a,Paduano2021}, and placed limits on the existence of intermediate mass BHs in GCs \citep{Tremou2018}. Accompanying the MAVERIC survey is a catalogue of over 1100 X-ray sources detected by the \chandra X-ray Observatory's ACIS detector from observations of 38 GCs \citep{Bahramian2020}.
    
    In this paper, we present the results of a multiwavelength study of exotic binaries in NGC 3201 using the MAVERIC survey and the MUSE GC survey. We identify 42 sources of various classes within the half-light radius of the cluster. We also present the first radio and X-ray limits on the detached BHs in NGC~3201. \S~\ref{sec:methods} details our catalogue selection and cross-matching taking into account coordinate uncertainties. \S~\ref{sec:results} presents the sources considered in this paper and possible interpretations of their natures. \S~\ref{sec:discussion} presents a discussion of these results, including an interpretation as to what the radio and X-ray limits on the detached BHs imply for the presence of accretion in these systems, how the number of XRBs in the cluster compares to other clusters, and an overview of the underluminous SSGs that are present in the cluster.

\section{Data and Reduction} \label{sec:methods}
    Our goal is to provide a comprehensive list of energetic sources within NGC 3201. In particular, we study all radio and X-ray sources in NGC~3201, and search for optical counterparts to these sources. We also investigate the radio and X-ray properties of some interesting sources discovered through the MUSE surveys of this cluster, namely the detached black holes and the SSG population in the cluster. To do this, we have combined data from various multiwavelength surveys to explore the full range of behaviours of sources within the cluster.

\subsection{Catalogue selection}
\subsubsection{The MAVERIC Survey}
    The MAVERIC survey contains a list of both radio and X-ray sources that are present in NGC~3201. NGC~3201 has been observed for a total of 18.1 hr with the ATCA, and catalogues of sources detected with significance of $>5\mhyphen\sigma$ and $>3\mhyphen\sigma$ were compiled (Tudor et al., in prep.). The $5\mhyphen\sigma$ catalogue represents radio sources that have a confident detection, and are the main radio sources we consider in this paper. The sources in the $3\mhyphen\sigma$ catalogue are only considered if they have a multiwavelength counterpart (e.g., optical or X-ray) that provides further evidence that there is actually a source present at that location. 
    
    The X-ray source catalogue of NGC~3201 contains 47 sources that are associated with NGC~3201 characterised by a detection quality parameter. This parameter indicates the confidence of the source detection. This parameter is discussed in depth in Section 4.2 in \citet{Bahramian2020}, and is calculated based on the minimum false probability value of the source and the source count rate. A minimum false probability value of $<1\%$ and a net count $\geq5$ is a confident detection (detection quality is 0), whereas a minimum false probability value $<1\%$ and a net count $<5$ is a marginal detection (detection quality is 1). For this paper, we consider all sources in the X-ray source catalogue for NGC~3201 with a detection quality of 0 or 1.
    
\subsubsection{Optical surveys}
    In this paper, optical surveys serve two purposes. Firstly, photometric surveys will allow us to construct colour-magnitude diagrams (CMDs) to investigate where a source lies relative to other sources in the cluster. Secondly, spectroscopic studies will allow us to search for radial velocity variations and identify sources in binary systems, allowing us to estimate the orbital parameters and the mass of a possibly invisible companion. 
    
    To investigate the photometric properties of the cluster, we used data from the HST UV Globular Cluster Survey \citep[HUGS, ][]{Piotto2015,Nardiello2018}. This survey includes photometric catalogues of 56 GCs and one open cluster in five photometric bands (F275W, F336W, F438W, F606W, and F814W). These data give us insight into the stellar populations in NGC 3201, and where various sources will fall on the CMD. 
    
    To investigate the spectroscopic properties of sources within the cluster, we used catalogues created by MUSE. NGC 3201 has been studied extensively with MUSE since the first discovery of a detached BH candidate within the cluster \citep{Giesers2018}. For our purposes we have used data from the binary and emission line catalogues of NGC~3201 produced by MUSE \citep{Giesers2019,Gottgens2019}. 
    
\subsection{X-ray data reduction}
    NGC 3201 has been observed once by \chandra  using ACIS-S in 2010 for \SI{85}{ks} under the observation ID 11031. To further investigate the X-ray properties of the cluster and perform X-ray spectroscopy beyond the scope of the analysis presented by \citet{Bahramian2020}, we downloaded and reduced this dataset. The analysis in \citet{Bahramian2020} included X-ray spectral fitting of each X-ray source with three spectral models. The models considered were a power-law emission model, an apec emission spectrum from ionised diffuse gas, and a blackbody radiation model.
    
    \textsc{ciao} 4.12.1 and \textsc{caldb} 4.9.3 \citep{Fruscione2006} were used to reduce and reprocess the data. The data were reprocessed using \verb#chandra_repro#, and source and background spectra were extracted using \verb#specextract#. Background and source regions varied depending on the position of the source in the X-ray image, and source crowding was not an issue. \textsc{xspec} 12.11 \citep{Arnaud1996} was used to perform X-ray spectral analysis. 
    
\subsection{Catalogue cross-matching and source identification}
    Prior to any catalogue cross-matching, we first restricted the MAVERIC sample to only consider sources within the half-light radius of NGC 3201 (\SI{3.1}{\arcminute}). Anything outside this radius was considered to have a higher chance of not being associated with the cluster, either as a foreground or background source. The MAVERIC survey listed 39 radio sources with a detection limit of $5\mhyphen\sigma$ associated with NGC 3201, of which 13 sources are within the half-light radius. The accompanying X-ray catalogue listed 24 X-ray sources with a detection quality of 0 or 1 within the half-light radius of the cluster.
    
    To identify counterparts across the catalogues and surveys we are considering, we initially searched for matches within \SI{1}{\arcsecond} of the input coordinates. This was a conservative threshold to account for the uncertainties associated with our radio and X-ray coordinates. The nominal uncertainty in the radio coordinates is $<\SI{0.1}{\arcsecond}$. The uncertainty in the X-ray coordinates is a combination of the overall frame alignment and statistical precision of the individual X-ray sources. Through cross-matching with the HUGS catalogue, we determined that the uncertainty in the frame alignment is $\sim\SI{0.3}{\arcsecond}$. The average uncertainty in the statistical precision of individual X-ray sources was determined by calculating the separation between the centroid X-ray coordinates (coordinates estimated through centroiding) and the reconstructed X-ray coordinates (coordinates estimated through X-ray image reconstruction\footnote{This is a standard method used in the software package \textsc{acis extract} \citep{Broos2010}.}), with the coordinates provided in \citet{Bahramian2020}. This value is $\sim\SI{0.3}{\arcsecond}$, giving an overall uncertainty in X-ray coordinates of $\sim\SI{0.42}{\arcsecond}$. For the optical/X-ray sources we discuss in \S~\ref{sec:optical_x-rays}, we calculate the individual X-ray source uncertainty for each potential cross-match. Any matches identified across multiple catalogues were also visually inspected to check the quality of the cross-match and identify any other possible candidate matches within errors. This also allowed us to reject any initial cross-matches between the catalogues with a difference between the input coordinates of $\gtrsim\SI{0.5}{\arcsecond}$.

\section{Results} \label{sec:results}
\begin{table*}
    \centering
        \caption{The 42 sources studied in this work, with the detached BHs listed first. The coordinates listed for the MUSE binary sources are taken from \citep{Giesers2019}. For the optical/X-ray sources and X-ray only sources, we take the X-ray coordinates from the MAVERIC X-ray source catalogue. For the remaining radio only sources, we give radio coordinates from the MAVERIC radio survey. The source type column indicates which survey the source is detected in. MUSE binaries are detected in the MUSE binary catalogue of NGC~3201 \citep{Giesers2019}, optical sources are detected in the HUGS survey \citep{Piotto2015,Nardiello2018}, X-ray sources are detected in the MAVERIC X-ray source catalogue \citep{Bahramian2020}, and radio sources are detected in the MAVERIC radio survey \citep[][Tudor et al., in prep.]{Shishkovsky2020}. The final column indicates the likely nature of each source.}
        \label{tab:all_sources}
    \begin{threeparttable}
        \begin{tabular}{lccll}
            \hline
            \hline
            Source                      & RA           & Dec & Source type   & Likely nature                              \\
            \hline
            ACS ID \#12560 \tnote{1}             & 10:17:37.090 & -46:24:55.332 & MUSE binary/detached BH    & Cluster member                       \\
            ACS ID \#21859 \tnote{2}             & 10:17:39.233 & -46:24:24.876 & MUSE binary/detached BH    & Cluster member                     \\
            ACS ID \#5132  \tnote{2}             & 10:17:36.082 & -46:25:33.060 & MUSE binary/detached BH    & Cluster member         \\
            CXOU J101737.58-462352.2 \tnote{2}   & 10:17:37.589 & -46:23:52.246 & MUSE binary/X-ray source   & Cluster member              \\
            CXOU J101735.57-462450.5 \tnote{2}   & 10:17:35.582 & -46:24:50.562 & MUSE binary/X-ray source   & Cluster member             \\
            ACS ID \#14749 \tnote{3}             & 10:17:33.146 & -46:25:07.428 & MUSE binary                & Cluster member             \\
            ACS ID \#11405 \tnote{2}             & 10:17:39.257 & -46:25:11.892 & MUSE binary                & Cluster member             \\
            CXOU J101730.49-462442.4    & 10:17:30.489 & -46:24:42.437          & Optical/X-ray source       & Cluster member                         \\
            CXOU J101727.83-462500.6    & 10:17:27.836 & -46:25:00.595          & Optical/X-ray source       & Background?                                               \\
            CXOU J101736.06-462422.5    & 10:17:36.070 & -46:24:22.619          & Optical/X-ray source       & Cluster member                                                \\
            CXOU J101735.79-462418.1    & 10:17:35.795 & -46:24:18.101          & Optical/X-ray source       & Cluster member                                                \\
            CXOU J101735.85-462346.1    & 10:17:35.842 & -46:23:46.064          & Optical/X-ray source       & Cluster member                                                \\
            CXOU J101729.85-462440.6    & 10:17:29.824 & -46:24:40.692          & Radio source/X-ray source  & Background                                                \\
            CXOU J101729.25-462644.0    & 10:17:29.259 & -46:26:43.954          & Radio source/X-ray source  & Background                                                \\
            CXOU J101736.17-462539.5    & 10:17:36.173 & -46:25:39.526          & X-ray source               & Background?                                                \\
            CXOU J101742.96-462509.1    & 10:17:42.939 & -46:25:09.426          & X-ray source               & Background?                                                \\
            CXOU J101737.30-462332.0    & 10:17:37.313 & -46:23:32.089          & X-ray source               & Background?                                                \\
            CXOU J101730.77-462348.2    & 10:17:30.757 & -46:23:48.286          & X-ray source               & Background?                                                \\
            CXOU J101741.33-462554.6    & 10:17:41.361 & -46:25:55.006          & X-ray source               & Background?                                                \\
            CXOU J101730.60-462555.2    & 10:17:30.619 & -46:25:55.121          & X-ray source               & Background?                                                \\
            CXOU J101725.45-462452.3    & 10:17:25.461 & -46:24:52.434          & X-ray source               & Background?                                                \\
            CXOU J101739.24-462242.6    & 10:17:39.246 & -46:22:42.625          & X-ray source               & Background?                                                \\
            CXOU J101726.64-462644.9    & 10:17:26.646 & -46:26:45.276          & X-ray source               & Background?                                                \\
            CXOU J101722.88-462334.0    & 10:17:22.886 & -46:23:33.994          & X-ray source               & Background?                                                \\
            CXOU J101730.66-462714.9    & 10:17:30.709 & -46:27:14.796          & X-ray source               & Background?                                                \\
            CXOU J101739.49-462200.7    & 10:17:39.485 & -46:22:00.617          & X-ray source               & Background?                                                \\
            CXOU J101723.71-462633.7    & 10:17:23.700 & -46:26:33.590          & X-ray source               & Background?                                                \\
            CXOU J101727.26-462214.2    & 10:17:27.251 & -46:22:14.315          & X-ray source               & Background?                                                \\
            CXOU J101749.73-462243.3    & 10:17:49.743 & -46:22:43.129          & X-ray source               & Background?                                                \\
            ATCA J101742.667-462454.262 & 10:17:42.667 & -46:24:54.262          & Radio source               & Background?                                                \\
            ATCA J101726.705-462504.558 & 10:17:26.705 & -46:25:04.558          & Radio source               & Background?                                                \\
            ATCA J101732.309-462626.163 & 10:17:32.309 & -46:26:26.163          & Radio source               & Background?                                                \\
            ATCA J101731.164-462642.881 & 10:17:31.164 & -46:26:42.881          & Radio source               & Background?                                                \\
            ATCA J101744.735-462631.964 & 10:17:44.735 & -46:26:31.964          & Radio source               & Background?                                                \\
            ATCA J101746.690-462306.033 & 10:17:46.690 & -46:23:06.033          & Radio source               & Background?                                                \\
            ATCA J101743.829-462236.644 & 10:17:43.829 & -46:22:36.644          & Radio source               & Background?                                                \\
            ATCA J101723.716-462322.616 & 10:17:23.716 & -46:23:22.616          & Radio source               & Background?                                                \\
            ATCA J101721.425-462536.169 & 10:17:21.425 & -46:25:36.169          & Radio source               & Background?                                                \\
            ATCA J101748.939-462245.159 & 10:17:48.939 & -46:22:45.159          & Radio source               & Background?                                                \\
            ATCA J101727.933-462712.667 & 10:17:27.933 & -46:27:12.667          & Radio source               & Background?                                                \\
            ATCA J101749.983-462254.064 & 10:17:49.983 & -46:22:54.064          & Radio source               & Background?                                                \\
            ATCA J101740.920-462144.955 & 10:17:40.920 & -46:21:44.955          & Radio source               & Background?                                                \\
            \hline
        \end{tabular}
        \begin{tablenotes}
            \item[1] \citet{Giesers2018}
            \item[2] \citet{Giesers2019}
            \item[3] \citet{Kaluzny2016,Giesers2019}
        \end{tablenotes}
    \end{threeparttable}
\end{table*}

    Through cross-matching the MAVERIC survey with the HUGS survey and the MUSE binary and emission line catalogue we identified two MUSE binary sources that have X-ray emission, five X-ray sources within the HUGS survey, and two radio/X-ray counterparts from within the MAVERIC survey. We also identify 15 X-ray sources and 13 radio sources within the half-light radius of NGC 3201. These sources are shown in Table~\ref{tab:all_sources}. 
    
    Throughout this paper, there are multiple times where upper limits on the radio and X-ray luminosities are calculated. For consistency, we used the same approach for each source. To calculate a $3\mhyphen\sigma$ upper limit on the 5.5 GHz radio flux density, we take three times the central RMS noise of the 5.5 GHz image. This gives a $3\mhyphen\sigma$ upper limit of \SI{11.7}{\micro\jansky} (Tudor et al., in prep.). We note that \citet{Tremou2018} takes a different value $3\mhyphen\sigma$ upper limit for the radio images of NGC~3201. The value used by \citet{Tremou2018} is the $3\mhyphen\sigma$ upper limit of the stacked 7.25 GHz radio images. To retain sensitivity to steep spectrum sources, we have instead chosen to use the 5.5 GHz image and catalogue.
    
    To calculate a 90\% upper limit on the 1-10 keV X-ray flux of a source, we first determined the source and background counts. Source counts are determined from a circular region of radius \SI{1.5}{\arcsecond} around the source coordinates, and background counts are determined using an annulus region of inner radius \SI{1.9}{\arcsecond} and outer radius \SI{12.3}{\arcsecond}. The background counts are normalised, and we then took the 90\% upper limit on the X-ray count rate using the method of \cite{Kraft1991}. This count rate upper limit was converted to a flux upper limit using the exposure time of the \chandra observation and modelling the emission with a power law with an index of $\Gamma=1.7$. The flux was converted to a luminosity upper limit using $4\pi d^2 F_{\textrm{x}}$. In all conversions to luminosity, the distance to NGC~3201 is assumed to be 4.74 kpc \citep{Baumgardt2019}.

\subsection{Known BHs} \label{sec:BH-1}

    \begin{table*}
        \centering
        \caption{The $3\mhyphen\sigma$ 5.5 GHz radio upper limits and the 90\% 1-10 keV X-ray upper limits of the two confirmed and one candidate BH in NGC~3201.}
        \label{tab:bh_limits}
        \begin{tabular}{lcccc}
            \hline
            \hline
            Source & 5.5 GHz flux density & 5.5 GHz $L_{\textrm{R}}$ & 1-10 keV X-ray flux & 1-10 keV $L_{\textrm{X}}$ \\
                   & (\si{\micro\jansky})       & (\si{\erg\per\second})   & (\si{\erg\per\second\per\centi\metre\squared}) & (\si{\erg\per\second}) \\ 
            \hline
            ACS ID \#12560 & $<11.7$ & $<\num{1.7e27}$ & $<\num{5.2e-16}$ & $<\num{1.4e30}$ \\
            ACS ID \#21859 & $<11.7$ & $<\num{1.7e27}$ & $<\num{3.2e-16}$ & $<\num{8.6e29}$ \\
            ACS ID \#5132  & $<11.7$ & $<\num{1.7e27}$ & $<\num{3.7e-16}$ & $<\num{9.9e29}$ \\
            \hline
        \end{tabular}
    \end{table*}
 
    \begin{figure*}
        \centering
        \includegraphics[scale=0.5]{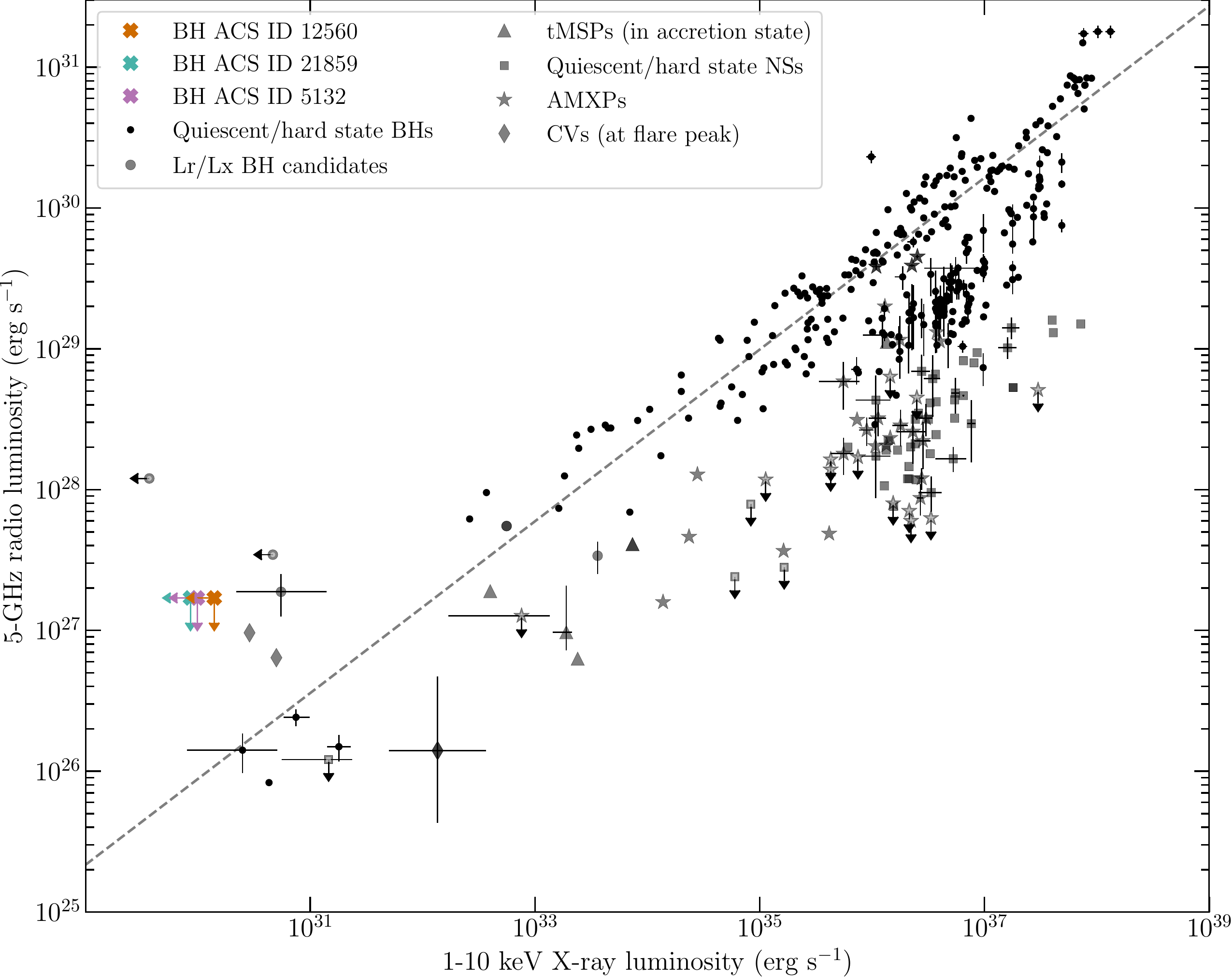}
        \caption{The 5 GHz radio and 1-10 keV X-ray luminosity limits for the BHs in NGC~3201 shown as crosses and indicated by their ACS ID numbers. The other points represent various other classes of accreting binary systems and are compiled from the database provided by \citet{Bahramian2018}.}
        \label{fig:lrlx}
    \end{figure*}   

    NGC 3201 contains two dynamically confirmed BHs and one candidate BH. Here we discuss the available radio and X-ray data from the MAVERIC survey for these systems, and calculate upper limits on the radio and X-ray emission of these sources. These limits are listed in Table~\ref{tab:bh_limits}, and Figure~\ref{fig:lrlx} shows these sources on the radio--X-ray luminosity plane. 
    
    All three sources have no radio counterpart at 5.5 GHz. We take the $3\mhyphen\sigma$ upper limit on the 5.5 GHz radio flux density to be \SI{11.7}{\micro\jansky} (three times the central RMS), which implies a $3\mhyphen\sigma$ 5.5 GHz radio luminosity upper limit of \SI{1.7e27}{\erg\per\second}. There is also no X-ray detection for any of the three sources, and we discuss the 90\% 1-10 keV X-ray upper limits individually for each source.
    
    ACS ID \#12560 was the first BH discovered in NGC~3201 by \citet{Giesers2018}. The source has a binary orbital period of $167.00\pm0.09$ days, with an eccentricity of $0.61\pm0.02$. The minimum mass of the BH is $4.53\pm0.21\ M_{\odot}$. The 90\% 1-10 keV X-ray flux upper limit is \SI{5.2e-16}{\erg\per\second\per\centi\metre\squared}, corresponding to a 90\% 1-10 keV X-ray luminosity upper limit of \SI{1.4e30}{\erg\per\second}.
    
    ACS ID \#21859 was discovered by \citet{Giesers2019}. The binary orbital period is $2.2422\pm0.0001$ days, much shorter than the other two BHs, and the orbital eccentricity is $0.07\pm0.04$. The minimum mass of the BH is $7.68\pm0.5\ M_{\odot}$. The 90\% upper limit on the 1-10 keV X-ray flux is \SI{3.2e-16}{\erg\per\second\per\centi\metre\squared}, giving a 90\% 1-10 keV X-ray luminosity upper limit of \SI{8.6e29}{\erg\per\second}.
    
    ACS ID \#5132 is a candidate BH discovered by \citet{Giesers2019}. The binary orbital period is $764\pm11$ days with an eccentricity of $0.28\pm0.16$. The minimum mass of the system is $4.40\pm2.82\ M_{\odot}$. The 90\% upper limit on the 1-10 keV X-ray flux of the source is \SI{3.7e-16}{\erg\per\second\per\centi\metre\squared}, implying a 90\% 1-10 keV X-ray luminosity upper limit of \SI{9.9e29}{\erg\per\second}.


\subsection{Sub-subgiants and red stragglers} \label{sec:results_ssgs}
    The MUSE binary catalogue \citep{Giesers2019} lists four SSGs within the field-of-view of the survey, identified based on positions of the sources on the cluster CMD and by fitting the radial velocity variations observed with a Keplerian orbit. Two of these sources are also emission line sources as noted in the emission line catalogue \citep{Gottgens2019}. These four sources are discussed in depth below. Additionally, we have discovered one new candidate red straggler system with X-ray emission, based on its position on the cluster CMD. Here we discuss the multiwavelength properties of these systems.

\subsubsection{CXOU J101730.49-462442.4}
    CXOU J101730.49-462442.4 is an X-ray source listed by the MAVERIC survey \citep{Shishkovsky2020, Bahramian2020} that is consistent with an optical source in the HUGS survey (source R0001757). The optical and X-ray coordinates are consistent to within $<0.1\si{\arcsecond}$. Based on its position on the CMD of NGC 3201 constructed from the HUGS F606W and F814W magnitudes (see Figure~\ref{fig:hugs_cmd_ssgs}), it is a red straggler system. We classify the source as a red straggler instead of a SSG as it is brighter than subgiants but lies redward of the giant branch. This source has a F606W magnitude of 16.8784, and a probability of 97.8\% of being a cluster member \citep{Piotto2015,Nardiello2018}. The source is not covered by current MUSE observations of the cluster. 
    
    From the MAVERIC survey \citep{Bahramian2020}, this source is best fit by an apec emission spectrum from ionised diffuse gas, with a power-law model having a relative probability of 0.694. We fit an absorbed apec model (\verb#tbabs#$\times$\verb#cflux#$\times$\verb#apec#) to the spectrum to calculate the 1-10 keV X-ray flux of the source. To estimate the cluster hydrogen column density, we use the $E(B-V)$ values estimated by \citet{Harris1996}, and the correlations estimated by \cite{Bahramian2015} and \citet{Foight2016}. This gives a hydrogen column density of $N_H=\SI{2e21}{\per\squared\centi\metre}$. By freezing the absorption parameter in our model to this value, the 1-10 keV X-ray flux is $5.8^{+2.8}_{-1.8}\times10^{-15}$ \si{\erg\per\second\per\centi\metre\squared} with an electron temperature of $kT=3.9^{+6.2}_{-1.6}$ \si{keV}. This implies a 1-10 keV luminosity of \SI{1.6e31}{\erg\per\second}. There is no radio detection of this source so we adopt a $3\mhyphen\sigma$ 5.5 GHz flux density upper limit of \SI{11.7}{\micro\jansky}, corresponding to a 5.5 GHz luminosity upper limit of \SI{1.7e27}{\erg\per\second}.

\subsubsection{CXOU J101737.58-462352.2/ACS ID \#22692}
    CXOU J101737.58-462352.2 is a SSG star detected by MUSE \citep{Giesers2019}. This system has an orbital period of $5.1038\pm0.0004$ days and a low eccentricity of $0.02\pm0.03$, with a F606W magnitude of 17.25. This source shows X-ray emission and varying H$\alpha$ emission lines \citep{Giesers2019,Gottgens2019}. 
    
    This source is detected as an X-ray source in the MAVERIC survey, and its X-ray spectrum is best fit with a blackbody radiation model. We fit an absorbed blackbody radiation model (\verb#tbabs#$\times$\verb#cflux#$\times$\verb#bbodyrad#) to the X-ray spectrum, with the hydrogen column density frozen to the cluster value, to calculate the 1-10 keV X-ray flux from the source. The 1-10 keV X-ray flux of the source is $1.1^{+0.6}_{-0.4}\times10^{-15}$ \si{\erg\per\second\per\centi\metre\squared}, corresponding to a 1-10 keV X-ray luminosity of \SI{3.0e30}{\erg\per\second}, and the electron temperature is $kT=0.3\pm0.1$ keV. This source has no radio counterpart, with a $3\mhyphen\sigma$ 5.5 GHz upper limit of \SI{11.7}{\micro\jansky}, implying a 5.5 GHz luminosity upper limit of \SI{1.7e27}{\erg\per\second}. The X-ray luminosity of this source is the highest for the SSGs detected by MUSE, which could be linked to its short orbital period.
    
    
    The minimum invisible mass of this system is $0.35\pm0.03\ M_{\odot}$. By assuming that the minimum invisible mass has Gaussian errors and by isotropically sampling the inclination angle between $\cos5\si{\degree}$ and $\cos90\si{\degree}$, we use a Monte Carlo simulation to estimate the most probable mass of the unseen companion. We use a lower bound on the inclination angle of \SI{5}{\degree} as for inclination angles smaller than this, no radial velocity variations would be observed. The unseen companion has a median mass of 0.408 $M_{\odot}$, with a 90\% confidence interval of 0.322 $M_{\odot}$ to 1.104 $M_{\odot}$. This indicates that the companion object in this system is either a white dwarf or another type of star, such as a M dwarf.

\subsubsection{CXOU J101735.57-462450.5/ACS ID \#13438}
    CXOU J101735.57-462450.5 is a SSG star detected by MUSE \citep{Giesers2019}. The orbital period of this system is $5.9348\pm0.0003$ days, and the eccentricity is the same as the SSG discussed above at $0.02\pm0.03$. The F606W magnitude of this system is 17.17, and the source shows X-ray emission with several MUSE spectra of this star showing a partially filled in H$\alpha$ absorption line \citep{Giesers2019,Gottgens2019}.
    
    This source is listed as an X-ray source in the MAVERIC survey, and is best fit by an apec model. By fitting an absorbed apec model to the X-ray spectrum of this source (and freezing the hydrogen column density to the cluster value), we calculate the 1-10 keV X-ray flux of this source to be $5.0^{+4.5}_{-2.8}\times10^{-16}$ \si{\erg\per\second\per\centi\metre\squared}, with an electron temperature of $kT=1.1^{+0.4}_{-0.3}$ \si{keV}. This X-ray flux implies a 1-10 keV X-ray luminosity of \SI{1.3e30}{\erg\per\second}. There is no radio counterpart to this source, with a $3\mhyphen\sigma$ 5.5 GHz upper limit of \SI{11.7}{\micro\jansky}, implying a 5.5 GHz luminosity upper limit of \SI{1.7e27}{\erg\per\second}.
    
    The minimum invisible mass of this system is the same as the above source, CXOU J101737.58-462352.2, at $0.35\pm0.03\ M_{\odot}$. A similar Monte Carlo simulation as described above indicate that the median mass of the invisible companion is 0.408 $M_{\odot}$, with a 90\% confidence interval of 0.322 $M_{\odot}$ to 1.104 $M_{\odot}$. Again, the companion object is either a white dwarf or another star.

\subsubsection{ACS ID \#14749} \label{sec:SSG-3}
    This SSG with ACS ID \#14749 is a known SSG star with a reported period of $10.0037\pm0.002$ days \citep{Kaluzny2016}. This source is detected by MUSE with a period of $10.006\pm0.002$ days, an eccentricity of $0.09\pm0.07$, and a F606W magnitude of 17.03. The source shows a partially filled in H$\alpha$ absorption line \citep{Giesers2019,Gottgens2019}. 
    
    This source is not a detected radio or X-ray source in the MAVERIC survey. The 3$\sigma$ 5.5 GHz radio upper limit is \SI{11.7}{\micro\jansky}. The 90\% upper limit for the 1-10 keV X-ray flux was calculated to be \SI{4.6e-16}{\erg\per\second\per\centi\metre\squared}.
    
    The minimum invisible mass of this system is $0.53\pm0.04\ M_{\odot}$. Monte Carlo simulations indicate that the median companion mass is 0.618 $M_{\odot}$, with a 90\% confidence interval of 0.493 $M_{\odot}$ to 1.648 $M_{\odot}$. The companion object in this system is either a white dwarf or another star, however if the system is more face-on, a NS companion becomes possible.

\subsubsection{ACS ID \#11405}
    This source (ACS ID \#11405) is a SSG star with a longer orbital period than the other three SSGs detected by MUSE, with an orbital period of $17.219\pm0.006$ days and a higher eccentricity of $0.42\pm0.08$. The F606W magnitude of this star is 17.25, and the source shows a partially filled in H$\alpha$ absorption line \citep{Giesers2019,Gottgens2019}.
    
    This source has no radio or X-ray counterpart in the MAVERIC survey. The 3$\sigma$ 5.5 GHz radio upper limit for this source is \SI{11.7}{\micro\jansky}. The 90\% 1-10 keV X-ray flux upper limit is \SI{5.7e-16}{\erg\per\second\per\centi\metre\squared}.
    
    The minimum invisible mass of this source is lower than that for the other SSGs, at $0.15\pm0.02\ M_{\odot}$. The median mass indicated by Monte Carlo simulations is $0.179\ M_{\odot}$, with a 90\% confidence interval of $0.128\ M_{\odot}$ to $0.476\ M_{\odot}$. Due to the low median mass, the companion is most likely another main sequence star or an extremely low-mass white dwarf.

\subsection{Optical/X-ray sources} \label{sec:optical_x-rays}
    \begin{figure*}
        \centering
        \includegraphics[scale=0.48]{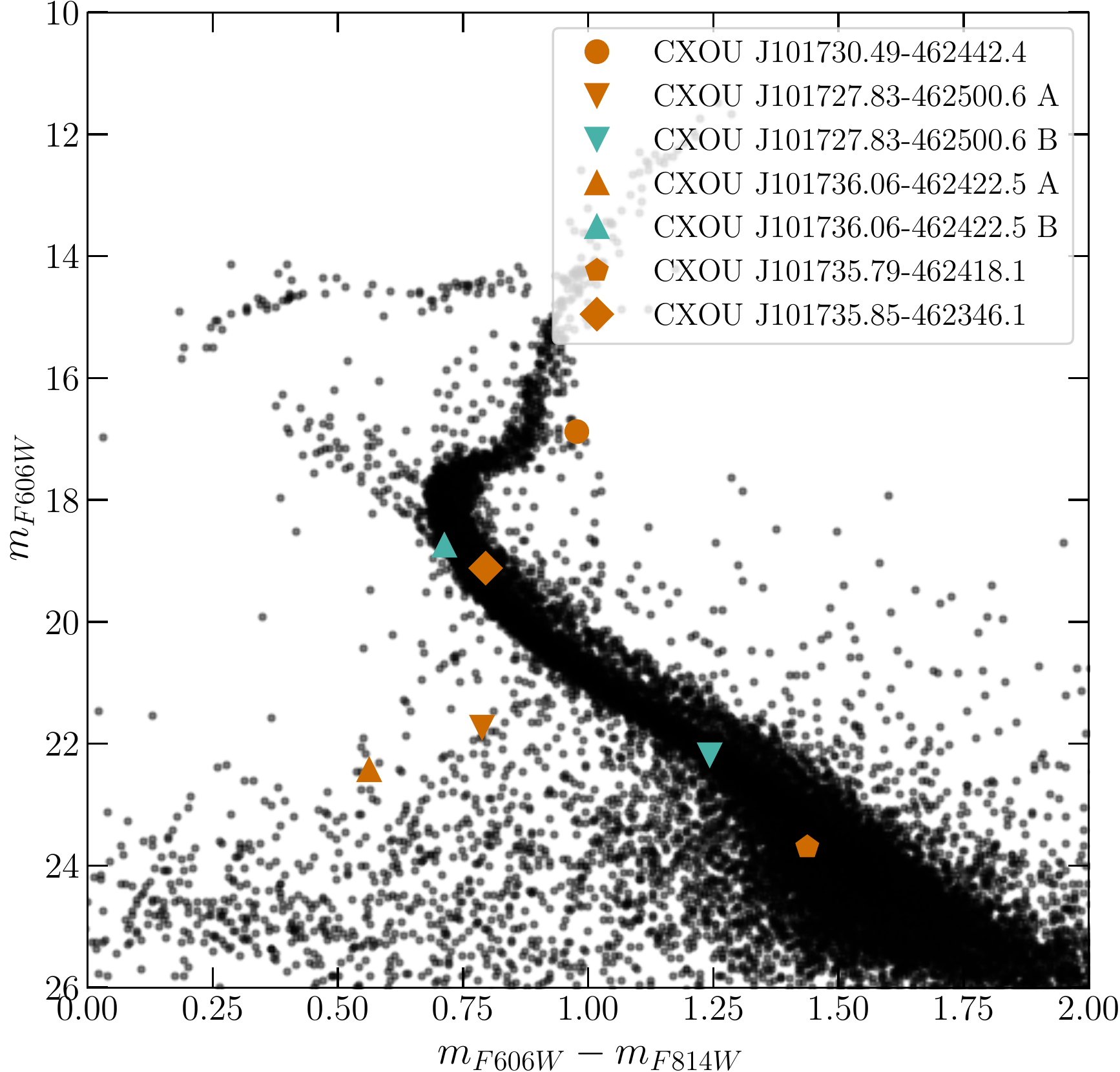}
        \includegraphics[scale=0.48]{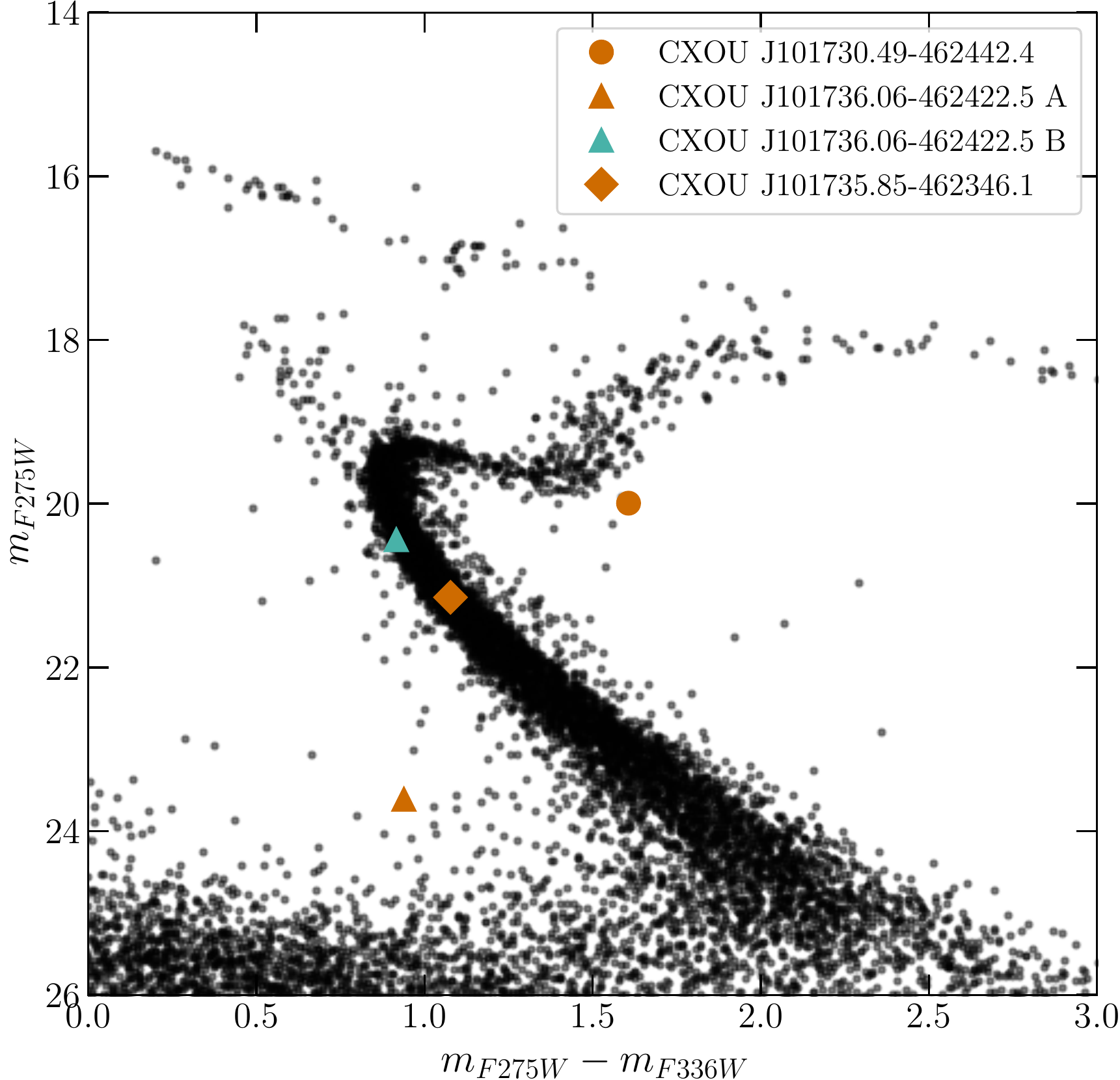}
        \caption{\textit{Left}: The visible CMD of NGC~3201 constructed using the F606W and F814W photometric bands. \textit{Right}: The UV CMD of NGC~3201 constructed using the F275W and F336W photometric bands. For both CMDs we indicate the most likely optical counterparts for the X-ray sources that have HUGS counterparts. Markers of the same shape with different colours are used to indicate where there is more than one possible optical counterpart for an X-ray source.}
        \label{fig:hugs_cmd_ssgs}
    \end{figure*}

    In addition to the candidate red straggler CXOU J101730.49-462442.4, there are four other HUGS sources that we identify X-ray counterparts for in the MAVERIC survey. Only two sources, CXOU J101736.06-462422.5 and CXOU J101735.79-462418.1, are covered by current MUSE observations of the cluster. Below we summarise the main results for each of these sources, and the position of each source on both the visible and UV CMD of the cluster, as shown in Figure~\ref{fig:hugs_cmd_ssgs}. The X-ray properties of these sources are also shown in Table~\ref{tab:xray_sources}. For these sources, we also discuss the active galactic nuclei (AGN) probability and the cluster membership probability where listed in the HUGS survey. The AGN probability is a parameter included in the MAVERIC X-ray source catalogue, and is the probability that a source is a background AGN based solely on the source flux and the source position within the cluster, and ignores all other source information. More details about this parameter are provided in Section 4.7 of \citet{Bahramian2020}.
    
\begin{figure*}
        \centering
        \includegraphics[scale=0.48]{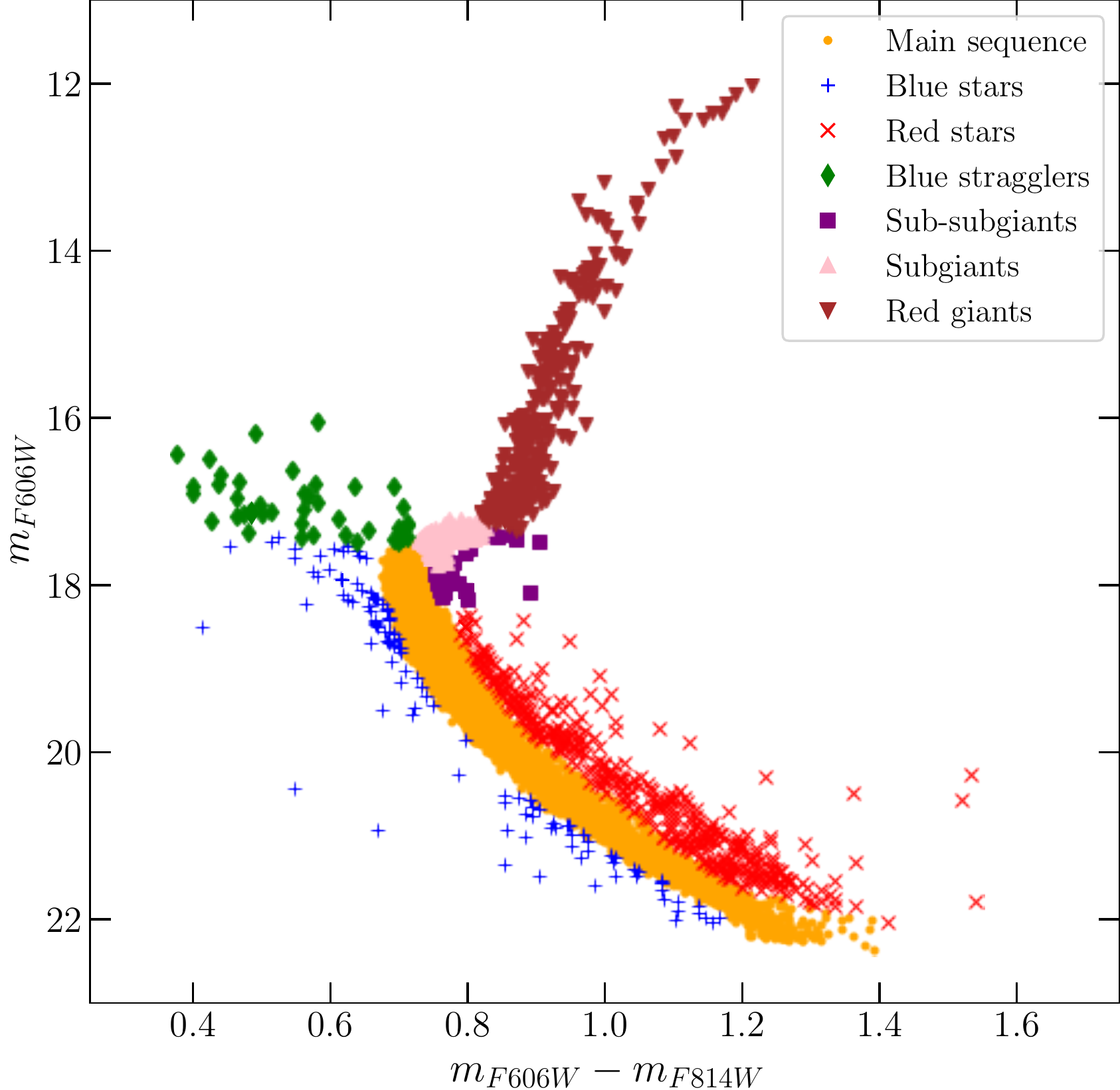}
        \includegraphics[scale=0.48]{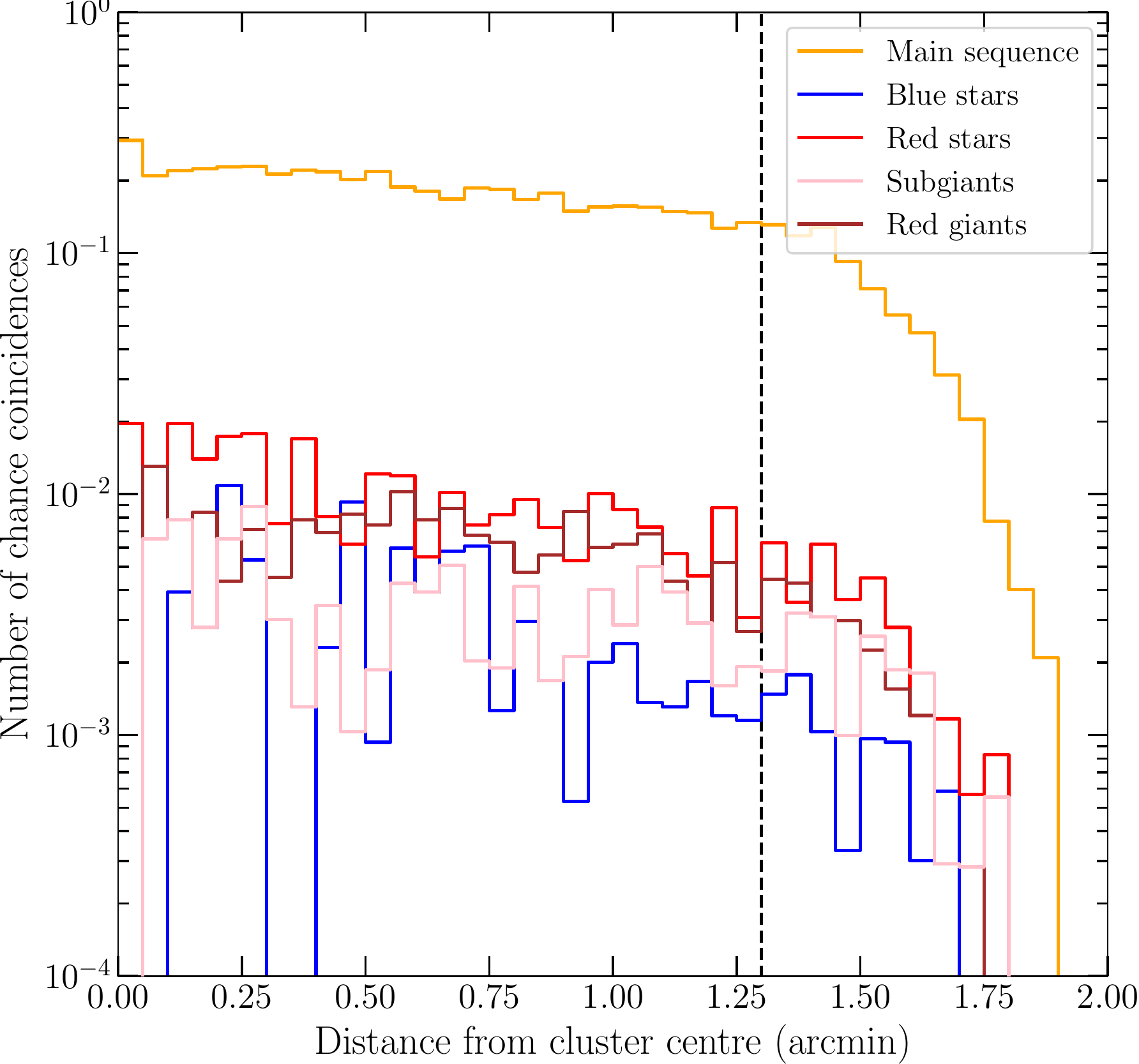}
        \caption{\textit{Left}: The visible CMD of NGC~3201 plotted for stars with a cluster membership probability $>0.9$. The different sub-populations are indicated with different markers. \textit{Right}: The number of chance coincidences expected for each sub-population plotted against the distance in arcmin from the cluster centre. The vertical dashed line indicates the core radius of NGC~3201, and the drop-off seen at $\sim\SI{2}{\arcmin}$ is due to the ACS field-of-view. The number of chance coincidences for main sequence stars is roughly constant within the cluster core, before decreasing beyond the core. Within the core, we expect roughly five optical sources to be a chance coincidence with an X-ray source.}
        \label{fig:cc_calcs}
    \end{figure*}

    Given the uncertainty of the X-ray coordinates, it is possible that any optical source with a position that is consistent with the X-ray source position could be a chance coincidence rather than the true counterpart. To estimate the number of chance coincidences, we follow the method outlined in Section 3.8.3 of \citet{Zhao2020a}. We use the visible CMD, plotted for stars that have a cluster membership probability of $>0.9$, to separate out the different stellar sub-populations in the cluster, shown in the left panel of Figure~\ref{fig:cc_calcs}. The sub-populations were separated using polygon selection areas using the \textsc{gluevis} software \citep{Beaumont2015,Robitaille2017}. The cluster was divided into several concentric annuli of radius \SI{0.05}{\arcminute}, and the numbers of chance coincidences for the different sub-populations were calculated for each annulus by assuming that each sub-population was evenly distributed within the annulus. The number of chance coincidences is given by
    \begin{equation}
        N_{\textrm{c}} = N_{\textrm{total}}\frac{\overline{A_{\textrm{err}}}}{A_{\textrm{annulus}}},
    \end{equation}
    where $N_{\textrm{c}}$ is the number of chance coincidences in a specific annulus, $N_{\textrm{total}}$ is the total number of stars in a sub-population for the specific annulus, $\overline{A_{\textrm{err}}}$ is roughly the uncertainty in the X-ray coordinates (\SI{0.42}{\arcsecond}), and $A_{\textrm{annulus}}$ is the area of the specific annulus. The number of chance coincidences within each annulus for each sub-population is shown in the right panel of Figure~\ref{fig:cc_calcs}. For main sequence stars, the number of chance coincidences is relatively constant within the core radius of the cluster, before sharply decreasing outside of the core, a trend also seen in the other sub-populations. This is consistent with what is expected for non-core-collapsed clusters, such as NGC~3201. The drop-off seen at a radius of $\sim\SI{2}{\arcmin}$ is due to the ACS field-of-view. Within the core radius of the cluster, we expect the following numbers of chance coincidences for each sub-population: $\approx4.91$ for main sequence stars, $\approx0.07$ for blue stars, $\approx0.27$ for red stars, $\approx0.09$ for sub-giant stars, and $\approx0.17$ for red giant stars. Due to the small numbers of stars within the SSG and blue straggler sub-populations, we calculate the number of chance coincidences over the entire ACS field-of-view ($\SI{202}{\arcsecond}\times\ \SI{202}{\arcsecond}$) instead. This gives the number of chance coincidences for SSGs and blue stragglers as $\approx\num{4.5e-4}$ and $\approx\num{6.0e-4}$ respectively.
    
    CXOU J101727.83-462500.6 is an X-ray source that falls outside the core radius of the cluster. The uncertainty on the X-ray coordinates is $\sim\SI{0.31}{\arcsecond}$. The X-ray spectrum of the source is best fit by an apec model, and has a 0.5-10 keV X-ray luminosity of $2.6^{+0.7}_{-0.6}\times10^{31}$ \si{\erg\per\second}. Within the \SI{0.31}{\arcsecond} uncertainty of the X-ray coordinates, there are two optical sources from the HUGS catalogue: R0015163, and R0015164. R0015163 falls on the main sequence on the cluster CMD (listed as CXOU J101727.83-462500.6 B), making a chance coincidence with a main sequence star likely. R0015164 lies blueward of the main sequence (listed as CXOU J101727.83-462500.6 A) with a 606W magnitude of 21.7276, making a chance coincidence less likely, and has a separation between optical and X-ray coordinates of $<\SI{0.1}{\arcsecond}$. Due to this, we favour R0015164 as the more likely counterpart. The AGN probability for this source is 1.0, and the HUGS survey has no listed cluster membership probability for this source. If the source is a cluster member, the X-ray luminosity and CMD position indicate that the source could be an XRB. However, given the lack of a secure optical counterpart and AGN probability, an AGN classification is the more feasible explanation.

    CXOU J101736.06-462422.5 is an X-ray source listed in the MAVERIC survey which falls well within the core of the cluster. The uncertainty on the X-ray coordinates is $\sim\SI{0.31}{\arcsecond}$. Its X-ray spectrum is best fit by an apec model, and has a 0.5-10 keV X-ray luminosity of $2.3^{+0.6}_{-0.5}\times10^{31}$ \si{\erg\per\second}. There are two optical sources from the HUGS survey within the X-ray uncertainty region: R0022094, and R0002240. R0022094 lies closer to the coordinates of the X-ray source ($\sim\SI{0.1}{\arcsecond}$), and lies blueward of the main sequence on the cluster CMD (listed as CXOU J101736.06-462422.5 A) with a F606W magnitude of 22.427, making a chance coincidence less likely. This source is too faint to be detected by MUSE. R0002240 lies $\sim\SI{0.18}{\arcsecond}$ from the X-ray source coordinates and is a likely radial velocity variable (probability $\sim0.7$), but a Keplerian orbit could not be fitted to the data through Monte Carlo simulations using the software \textsc{the joker} \citep{Price-Whelan2017}. This source sits on the main sequence of the cluster CMD (listed as CXOU J101736.06-462422.5 B), with a F606W magnitude of 18.7414 and a optical cluster membership probability of 0.971. Due to it falling on the main sequence, a chance coincidence is more likely for this source. However, given the radial velocity variations of this source, we favour R0002240 as the most likely optical counterpart to this X-ray source. This, along with its X-ray luminosity, indicate it could be an XRB or a CV.
    CXOU J101735.79-462418.1 is an X-ray source within the core of the cluster. The uncertainty on the X-ray coordinates is $\sim\SI{0.31}{\arcsecond}$. The X-ray spectrum of this source is best fit by a blackbody radiation model with a 0.5-10 keV X-ray luminosity of $4.0^{+0.8}_{-0.7}\times10^{31}$ \si{\erg\per\second}. Only one optical source falls within the X-ray uncertainty region, R0022891. R0022891 is $\sim\SI{0.1}{\arcsecond}$ from the X-ray coordinates, has a F606W magnitude of 23.6928, and lies towards the lower end of the main sequence on the cluster CMD, meaning it could be a chance coincidence. The AGN probability of the X-ray source is 0.43. If R0022891 is the optical counterpart, then the X-ray luminosity and CMD position indicate it could either be an XRB or a CV.

    CXOU J101735.85-462346.1 is an X-ray source within $\sim\SI{0.16}{\arcsecond}$ of the HUGS source R0027430. This source lies within the core of the cluster. The uncertainty on the X-ray coordinates is $\sim\SI{0.51}{\arcsecond}$. The X-ray spectrum of the source is best fit with a power-law model, however further inspection of this fit indicates that it is not physical due to its shallow photon index. Given that we cannot distinguish between an apec and a blackbody radiation model, we default to a blackbody radiation model due to it being simpler. A blackbody model fit gives a 0.5-10 keV X-ray luminosity $8.3^{+89.0}_{-8.1}\times10^{30}$ \si{\erg\per\second}. The AGN probability of this source is listed as 0.39. This source lies on the main sequence of the cluster CMD with a F606W magnitude of 19.1185, meaning a chance coincidence is possible, and has an optical cluster membership probability of 0.978. If R0027430 is the optical counterpart, then the X-ray luminosity and CMD position indicate that an AB, CV, or XRB classification is feasible.
    
    In addition to these four sources, there are a further three X-ray sources that have possible optical counterparts within the X-ray uncertainty regions. These X-ray sources are CXOU J101737.30-462332.0, CXOU J101725.45-462452.3, and CXOU J101730.77-462348.2. Upon visual inspection of these cross-matches, we find that it is unlikely that these optical sources are counterparts to the X-ray sources as they fall towards the edges of the respective X-ray uncertainty regions. Furthermore, an AGN classification is more feasible for CXOU J101725.45-462452.3 and CXOU J101730.77-462348.2 based on the the AGN probability of the X-ray sources and the lack of an optical cluster membership probability for the optical sources.

\subsection{Interesting radio sources} \label{sec:radio_sources}
    By comparing the radio and X-ray source catalogues in the MAVERIC survey, we identified two sources that have both radio and X-ray emission. Here we discuss the properties of each source in order to draw some conclusions about the source class. 
    The radio spectral index ($S_{\nu}\propto\nu^{\alpha}$) values and errors of these sources have been calculated in the initial MAVERIC survey (Tudor et al., in prep.).
    
    CXOU J101729.85-462440.6 has no optical counterpart within 0.5" of its radio or X-ray coordinates, and lies within the core of the cluster at a distance of 1.20' from the centre of the cluster. The X-ray spectrum of this source is best fit by a power-law model with a 0.5-10 keV X-ray flux of $4.3^{+13.4}_{-2.9}\times10^{-15}$ \si{\erg\per\second\per\centi\metre\squared}, implying a 0.5-10 keV X-ray luminosity of $1.2^{+3.8}_{-0.8}\times10^{31}$ \si{\erg\per\second} and a power-law index of $\Gamma=2.2^{+1.5}_{-2.1}$. There is no spectral information below 1 keV so we do not have a good measure of how soft this source is. Either it is soft with a high absorption component, or hard with a low absorption component. The 5.5 GHz radio flux density of the source is $19.7\pm3.9$ \si{\micro\jansky}, corresponding to a 5.5 GHz radio luminosity of $(2.9\pm0.6)\times10^{27}\si{\erg\per\second}$, and the spectral index for this source is $\alpha<0.59\pm0.40$ . The AGN probability for the X-ray source is 0.17, however the combined radio/X-ray luminosity ratio places it well above what is expected for accreting BHs \citep[$L_{\textrm{R}} \propto L_{\textrm{X}}^{0.6}$;][]{Gallo2014} and NSs. This, in addition to the lack of an optical counterpart to the source, makes it likely that this source is an AGN. 
    
    CXOU J101729.25-462644.0 is another radio/X-ray source identified in the MAVERIC survey with no optical counterpart within 1". This source is outside the core of the cluster and is 2.37' from the centre of the cluster. The X-ray spectrum of this source is best fit with an apec model with a 0.5-10 keV flux of $(1.5\pm0.3)\times10^{-14}$ \si{\erg\per\second\per\centi\metre\squared}, implying a 0.5-10 keV X-ray luminosity of $4.2^{+0.9}_{-0.7}\times10^{31}$ \si{\erg\per\second}. The source has a 5.5 GHz radio flux density of $133\pm4$ \si{\micro\jansky}, corresponding to a 5.5 GHz radio luminosity of $(2.0\pm 0.1)\times10^{28}$ \si{\erg\per\second}, and the spectral index is $\alpha=-1.32\pm0.19$. Like the above source, it has a much higher radio/X-ray luminosity ratio than is expected from typical accreting systems, and the AGN probability of the X-ray source is 1.0. Combined with the lack of an optical counterpart, we conclude that this source is also most likely an AGN. 
    
    We also identify one radio source within the core radius of the cluster, ATCA J101742.667-462454.262. This source has a 5.5 GHz radio flux density of $25.4\pm3.7$ \si{\micro Jy} corresponding to a radio luminosity of $(3.8\pm0.5)\times10^{27}$ \si{\erg\per\second}, and has a spectral index of $\alpha=-0.87\pm0.61$. There is no X-ray detection of this source. The 90\% upper limit on the 1-10 keV X-ray flux is \SI{7.4e-16}{\erg\per\second\per\centi\metre\squared}, corresponding to a 90\% 1-10 keV X-ray luminosity upper limit of \SI{2.0e30}{\erg\per\second}. We speculate that an AGN classification of this source is also plausible based on its radio spectrum.

\subsection{X-ray sources} \label{sec:xray_sources}
    The MAVERIC survey lists 24 X-ray sources within the half-light radius of NGC~3201 that have a detection quality of 0 or 1. Of these 24 sources, two are known sub-subgiants in the cluster, five have optical counterparts, and two have radio counterparts.
    We list the remaining unclassified 15 sources with no multiwavelength counterparts in Table~\ref{tab:xray_sources} and discuss them below.
    
    Ten of these 15 sources have X-ray spectra that are best fit by power-law models. Four sources have spectra best fit by apec models, and the spectrum of CXOU J101739.49-462200.7 is best fit by a blackbody model. However, for all sources we cannot distinguish between the best-fitting model and second best-fitting model to $>99\%$ confidence. All sources are within the half light radius of the cluster (3.1'), with only three sources, CXOU J101736.17-462539, CXOU J101742.96-462509, and CXOU J101737.30-462348.2, within the core (1.3'). The 0.5-10 keV X-ray luminosity of these sources falls in the range of $10^{29-31}$ \si{\erg\per\second}, making them very faint X-ray sources. 
    
    We use the model from \citet{Georgakakis2008} to calculate the X-ray source counts in the 0.5-10 keV range. Using the 90\% upper limit calculated for the BH ACD ID \#12560 (see \S~\ref{sec:BH-1}) as a conservative upper limit on our sensitivity, we expect approximately 28 extra-galactic X-ray sources within an area of the sky the size of the half-light radius of NGC~3201. This means that it is possible that all the X-ray sources we observe can be attributed to background sources. However, as shown in \S~\ref{sec:optical_x-rays}, there are some X-ray sources that are confidently cluster members due to their optical counterparts. Thus, it is more likely that these 15 X-ray sources with no multiwavelength counterparts are background sources. However, if some sources are associated with the cluster, it is plausible that most of these sources are ABs due to their low X-ray luminosities. Only the brightest of these sources could potentially be CVs or XRBs.

    \begin{table*}
        \centering
        \caption{The list of X-ray sources discussed. This list contains the SSGs and candidate RS source discussed in \S~\ref{sec:results_ssgs}, the four optical/X-ray sources discussed in \S~\ref{sec:optical_x-rays}, and the 15 X-ray sources with no multiwavelength counterparts discussed in \S~\ref{sec:xray_sources}. For each source we list the distance from the centre of the cluster as a fraction of the core and half light radii, the AGN probability of the source (\S~\ref{sec:xray_sources}), and a variety of X-ray spectral model parameters. For a power-law (pl) model we list the photon index ($\Gamma$), for an apec model we list the plasma temperature (kT), and for a blackbody radiation (bbr) model we list the electron temperature (kT). For some sources (e.g. CXOU J101736.17-462539.5 and CXOU J101741.33-462554.6) the best-fit model parameters are not physical. We also list the X-ray luminosity for the best-fitting model for each source.}
        \label{tab:xray_sources}
        \begin{tabular}{llcccccclcl}
            \hline
            \hline
            Counterpart & Source                   & R/R$_{\textrm{c}}$ & R/R$_{\textrm{h}}$  & AGN prob. & pl $\Gamma$ & apec kT & bbr kT & Best-fit model & Best-fit $L_{\textrm{X}}$ \\
                        &             &                    &        &             &             & (keV)   & (keV)  &        & (\si{\erg\per\second})    \\
            \hline
            MUSE/X-ray & CXOU J101737.58-462352.2 & 0.68 & 0.29 & 0.19 & $2.4^{+1.2}_{-0.8}$ & $3.3^{+8.9}_{-1.6}$ & $0.3\pm0.1$ & bbr & $3.0^{+1.6}_{-1.1}\times10^{30}$\\
                       & CXOU J101735.57-462450.5 & 0.18 & 0.08 & 0.22 & $3.0^{+0.7}_{-0.8}$ & $1.1^{+0.4}_{-0.3}$ & $0.2\pm0.1$ & apec& $1.3^{+1.2}_{-0.7}\times10^{30}$\\
            \hline
            Optical/X-ray & CXOU J101730.49-462442.4 & 0.84 & 0.35 & 0.25 & $1.9^{+0.7}_{-0.5}$& $3.9^{+6.2}_{-1.6}$  & $0.6\pm0.1$          & apec & $1.6^{+0.8}_{-0.5}\times10^{31}$ \\
                          & CXOU J101727.83-462500.6 & 1.21 & 0.51 & 1.00 & $1.9\pm0.6$        & $7.5^{+13.4}_{-3.9}$ & $0.6\pm0.1$          & apec & $2.6^{+0.7}_{-0.6}\times10^{31}$  \\
                          & CXOU J101736.06-462422.5 & 0.30 & 0.13 & 0.31 & $2.2\pm0.7$        & $4.7^{+5.2}_{-1.7}$  & $0.5\pm0.1$          & apec & $2.3^{+0.6}_{-0.5}\times10^{31}$  \\
                          & CXOU J101735.79-462418.1 & 0.37 & 0.15 & 0.43 & $1.2\pm0.4$        & $15.3^{+10.7}_{-7.8}$  & $0.9\pm0.1$          & bbr & $4.0^{+0.8}_{-0.7}\times10^{31}$  \\
                          & CXOU J101735.85-462346.1 & 0.76 & 0.32 & 0.39 & $0.2^{+2.6}_{-1.0}$& $3.5^{+15.7}_{-3.4}$  & $3.6\pm^{+4.8}_{-3.6}$& bbr & $8.3^{+89.0}_{-8.1}\times10^{30}$ \\
            \hline
            Radio/X-ray   & CXOU J101729.85-462440.6 & 0.92 & 0.39 & 0.17 & $2.2^{+1.5}_{-2.1}$& $2.2^{+15.1}_{-1.7}$  & $0.9^{+1.6}_{-0.5}$  & pl   & $1.2^{+3.8}_{-0.8}\times10^{31}$ \\
                          & CXOU J101729.25-462644.0 & 1.82 & 0.76 & 1.00 & $1.2^{+0.4}_{-0.3}$& $12.4^{+12.6}_{-6.5}$  & $0.8\pm0.1$  & apec & $4.2^{+0.9}_{-0.7}\times10^{31}$ \\
            \hline
            X-ray & CXOU J101736.17-462539.5 & 0.71 & 0.30 & 0.56 & $0.1\pm0.6$           & $13.5^{+12.5}_{-8.9}$ & $2.5^{+3.6}_{-1.1}$ & pl   & $2.6^{+1.6}_{-1.1}\times10^{31}$\\
                  & CXOU J101742.96-462509.1 & 0.87 & 0.36 & 0.18 & $1.4^{+1.6}_{-1.3}$   & $10.6^{+13.5}_{-7.3}$ & $1.2^{+1.2}_{-0.5}$ & pl   & $1.7^{+1.8}_{-0.7}\times10^{31}$  \\
                  & CXOU J101737.30-462332.0 & 0.94 & 0.39 & 0.21 & $1.7^{+1.1}_{-0.8}$   & $6.8^{+14.4}_{-4.3}$  & $0.2^{+0.2}_{-0.1}$ & pl   & $1.8^{+0.9}_{-0.6}\times10^{31}$  \\
                  & CXOU J101730.77-462348.2 & 1.08 & 0.45 & 0.65 & $2.6^{+1.0}_{-0.9}$   & $2.1^{+8.9}_{-1.5}$   & $0.3^{+0.1}_{-0.1}$ & pl   & $2.8^{+2.7}_{-1.3}\times10^{30}$  \\
                  & CXOU J101741.33-462554.6 & 1.08 & 0.45 & 1.00 & $-0.1^{+2.5}_{-0.8}$ & $8.6^{+14.6}_{-6.5}$  & $4.7^{+4.0}_{-3.0}$ & pl   & $3.8^{+17.2}_{-2.0}\times10^{31}$ \\
                  & CXOU J101730.60-462555.2 & 1.22 & 0.51 & 0.76 & $1.8^{+1.3}_{-1.2}$  & $8.7^{+14.7}_{-6.1}$  & $0.9^{+0.3}_{-0.2}$ & pl   & $1.7^{+1.9}_{-0.7}\times10^{31}$  \\
                  & CXOU J101725.45-462452.3 & 1.51 & 0.63 & 0.72 & $1.2^{+1.2}_{-0.8}$  & $8.2^{+15.2}_{-6.3}$  & $0.7^{+0.6}_{-0.3}$ & apec & $3.4^{+2.6}_{-1.5}\times10^{30}$  \\
                  & CXOU J101739.24-462242.6 & 1.60 & 0.67 & 0.84 & $1.2^{+2.1}_{-1.7}$  & $5.2^{+14.7}_{-3.6}$  & $1.9^{+3.7}_{-1.1}$ & pl   & $1.7^{+5.1}_{-1.0}\times10^{31}$  \\
                  & CXOU J101726.64-462644.9 & 2.05 & 0.86 & 0.64 & $2.6^{+0.9}_{-0.6}$  & $4.1^{+8.1}_{-1.9}$   & $0.3\pm0.1$         & pl   & $6.2^{+3.5}_{-2.0}\times10^{30}$  \\
                  & CXOU J101722.88-462334.0 & 2.06 & 0.86 & 0.68 & $1.9^{+1.2}_{-0.8}$  & $3.9^{+12.3}_{-2.3}$  & $0.6\pm0.2$         & apec & $5.1^{+3.1}_{-2.2}\times10^{30}$  \\
                  & CXOU J101730.66-462714.9 & 2.09 & 0.88 & 0.75 & $3.1^{+0.8}_{-1.3}$  & $1.7^{+3.7}_{-1.0}$   & $0.5\pm0.2$         & apec & $1.3^{+5.4}_{-0.7}\times10^{31}$  \\
                  & CXOU J101739.49-462200.7 & 2.13 & 0.75 & 0.92 & $2.1^{+1.1}_{-1.0}$  & $6.5^{+14.0}_{-4.1}$  & $0.9^{+0.3}_{-0.2}$ & bbr  & $1.3^{+0.6}_{-0.4}\times10^{31}$  \\
                  & CXOU J101723.71-462633.7 & 2.23 & 0.94 & 1.00 & $1.8^{+1.7}_{-1.9}$  & $2.8^{+15.6}_{-2.0}$  & $1.5^{+2.1}_{-0.7}$ & pl   & $2.5^{+10.8}_{-1.3}\times10^{31}$ \\
                  & CXOU J101727.26-462214.2 & 2.31 & 0.97 & 0.61 & $2.2^{+1.4}_{-2.0}$  & $1.2^{+8.4}_{-1.2}$   & $0.2^{+1.6}_{-0.1}$ & apec & $5.4^{+36940.0}_{-4.4}\times10^{29}$ \\
                  & CXOU J101749.73-462243.3 & 2.32 & 0.97 & 0.59 & $1.3^{+1.8}_{-1.2}$  & $5.2^{+15.5}_{-4.2}$  & $0.6^{+1.5}_{-0.4}$ & pl   & $5.9^{+9.0}_{-3.9}\times10^{30}$  \\
            \hline
        \end{tabular}
    \end{table*}

\subsection{Other radio sources} \label{sec:pure_radio_sources}
    The MAVERIC survey lists 14 radio sources detected with a significance of $>5\mhyphen\sigma$ within the half-light radius of NGC~3201. Of these, one source is also an X-ray source (CXOU J101729.25-462644.0). The remaining 13 sources are listed in Table~\ref{tab:radio_sources}. The source ATCA J101742.667-462454.262 is discussed in \S~\ref{sec:radio_sources}, but is also included in Table~\ref{tab:radio_sources}.
    
    The other radio sources in the cluster have 5.5 GHz flux densities of order \SI{10}{\micro Jy} with the exception of ATCA J101748.939-462245.159, which has a flux density of $(182\pm2)$ \si{\micro Jy}. The spectral indices of these sources are consistent with being flat or negative, with only one source, ATCA J101723.716-462322.616, having an inverted spectrum of $\alpha<1.16 \pm 0.41$, and these sources all fall at least \SI{0.04}{\arcminute} outside the core of the cluster.
    
    To estimate the number of background sources we expect within the area of sky contained within the half-light radius of NGC~3201, we use the simulated source counts of \citet{Wilman2008}. For the half-light radius of \SI{3.1}{\arcminute}, we expect 19 background sources. This indicates that every radio source detected in NGC~3201 can be explained as a background source. 
    
    \begin{table*}
        \centering
        \caption{The list of radio sources discussed, which includes the radio/X-ray sources discussed in \S~\ref{sec:radio_sources}, and the 15 radio sources with no multiwavelength counterpart discussed in \S~\ref{sec:pure_radio_sources}. For each source we list the distance from the centre of the cluster as a fraction of the core and half-light radii, the 5.5 GHz radio flux density and the spectral index.}
        \label{tab:radio_sources}
        \begin{tabular}{llcccc}
            \hline
            \hline
            Counterpart & Source                   & R/R$_{\textrm{c}}$ & R/R$_{\textrm{h}}$ & 5.5 GHz flux density (\si{\micro Jy}) & Spectral index \\
            \hline
            Radio/X-ray & CXOU J101729.85-462440.6 & 0.92 & 0.39 & $19.7 \pm 3.9$ & $<0.59\pm0.40$ \\
                        & CXOU J101729.25-462644.0 & 1.82 & 0.76 & $133 \pm 4.0 $ & $-1.32\pm0.19$ \\
            \hline
            Radio & ATCA J101742.667-462454.262 & 0.78 & 0.33 & $25.4 \pm 3.7$ & $-0.87 \pm 0.61$  \\
                  & ATCA J101726.705-462504.558 & 1.36 & 0.57 & $26.6 \pm 4.0$ & $<-0.13 \pm 0.31$ \\
                  & ATCA J101732.309-462626.163 & 1.43 & 0.60 & $16.6 \pm 3.9$ & $0.47  \pm 0.70 $  \\
                  & ATCA J101731.164-462642.881 & 1.68 & 0.71 & $26.5 \pm 4.2$ & $<0.22 \pm 0.32$  \\
                  & ATCA J101744.735-462631.964 & 1.73 & 0.73 & $24.6 \pm 4.5$ & $-0.15 \pm 0.64$  \\
                  & ATCA J101746.690-462306.033 & 1.82 & 0.76 & $52.0 \pm 4.5$ & $-0.48 \pm 0.41$  \\
                  & ATCA J101743.829-462236.644 & 1.89 & 0.79 & $28.7 \pm 4.5$ & $<0.60 \pm 0.32$  \\
                  & ATCA J101723.716-462322.616 & 2.03 & 0.85 & $24.9 \pm 5.0$ & $<1.16 \pm 0.41$  \\
                  & ATCA J101721.425-462536.169 & 2.15 & 0.90 & $30.1 \pm 4.8$ & $<-0.15 \pm 0.33$ \\
                  & ATCA J101748.939-462245.159 & 2.22 & 0.93 & $182 \pm 2.0$    & $-0.30 \pm 0.41$  \\
                  & ATCA J101727.933-462712.667 & 2.23 & 0.94 & $29.2 \pm 4.9$ & $<0.65 \pm 0.35$  \\
                  & ATCA J101749.983-462254.064 & 2.25 & 0.95 & $25.0 \pm 5.5$ & $<0.74 \pm 0.45$  \\
                  & ATCA J101740.920-462144.955 & 2.37 & 0.99 & $43.6 \pm 5.5$ & $<0.29 \pm 0.26$  \\
            \hline
        \end{tabular}
    \end{table*}

\section{Discussion} \label{sec:discussion}
    Our main results are the non-detections of the two confirmed and one candidate BH in the GC NGC~3201. We report the $3\mhyphen\sigma$ radio luminosity upper limits at 5.5 GHz and the 90\% 1-10 keV X-ray luminosity radio upper limits for these three sources. We list these values in Table~\ref{tab:bh_limits} and discuss the implications of these results in \S~\ref{sec:acc_limits}. 
    
    We also present a comprehensive list of multiwavelength sources within the half-light radius of NGC~3201. We identify two MUSE binary sources with X-ray emission, five optical sources with X-ray emission, two sources displaying both radio and X-ray emission, 15 X-ray sources with no other multiwavelength counterpart, and 13 radio sources with no other multiwavelength counterpart. The X-ray sources present in the cluster allow us to make some estimates as to the total population of XRBs in NGC~3201 (\S~\ref{sec:xrbs}), and the detections of two known SSGs in the cluster allow us to briefly discuss the evolution of this class of object (\S~\ref{sec:ssg_disc}).
    
\subsection{Accretion limits on the detached black holes} \label{sec:acc_limits}
\subsubsection{Calculating the mass accretion limits from observations}
    Our upper limits on the radio and X-ray luminosities of the detached BHs in NGC~3201 (Table~\ref{tab:bh_limits}) allow us to constrain the mass accretion rates onto these sources. Using these upper limits, we can calculate the maximum mass accretion rate that would be visible in both the radio and X-ray bands. We calculate these limits both directly from the X-ray luminosity and by using the fundamental plane of BH activity \citep{Merloni2003,Falcke2004}.
    
    The X-ray luminosity $L_{\textrm{X}}$ is directly related to the mass accretion rate $\dot{M}$ through the standard equation
    \begin{equation}
        L_{\textrm{X}} = \epsilon \dot{M}c^2,
    \end{equation}
    where $\epsilon$ is the radiative efficiency of the accretion and $c$ is the speed of light. Thus for a given X-ray luminosity, we can constrain the observable product $\epsilon\dot{M}$ to which we would be sensitive.
    
    The radio luminosity $L_{\textrm{R}}$ allows us to provide a constraint on the mass accretion rate through the fundamental plane of BH activity. The fundamental plane describes the relation between the X-ray luminosity, radio luminosity, and the mass of hard state accreting BHs. This relationship spans several orders of magnitude in mass, ranging from stellar-mass BHs to supermassive BHs at the centres of galaxies \citep{Merloni2003,Falcke2004,Plotkin2012}. For this work, we use the following form of the fundamental plane \citep{Miller-Jones2012,Plotkin2012}
    \begin{equation}
        \log L_{\textrm{X}} = 1.44\log L_{\textrm{R}} - 0.89\log M - 5.95,
    \end{equation}
    where $L_{\textrm{X}}$ and $L_{\textrm{R}}$ are in \si{\erg\per\second} and the BH mass $M$ is in $M_{\odot}$. By substituting in the above relation for X-ray luminosity, we can again constrain the observable product $\epsilon\dot{M}$ to which we would be sensitive for a given radio luminosity and BH mass.
    
    By using the radio and X-ray luminosity upper limits shown in Table~\ref{tab:bh_limits} and the minimum BH masses given in \citet{Giesers2019}, we can provide two constraints on the product $\epsilon\dot{M}$ that would be detectable: one from the X-ray luminosity upper limits, and one from the radio luminosity upper limits and BH masses. These constraints are shown in Table~\ref{tab:mdot_limits}, and they represent the deepest such accretion limits onto a BH to date for a GC. The limits on $\epsilon\dot{M}$ derived from the X-ray luminosity are two orders of magnitude deeper than those derived from the fundamental plane.
    
    \begin{table}
        \centering
        \caption{The limits on the product $\epsilon\dot{M}$, derived from X-ray luminosity limits and from the fundamental plane for BHs respectively. Our X-ray observations allow us to probe $\epsilon\dot{M}$ values two orders of magnitude deeper than our radio observations allow.}
        \label{tab:mdot_limits}
        \begin{tabular}{lcc}
            \hline
            \hline
            Source         & $\epsilon\dot{M}$ from $L_{\textrm{X}}$ & $\epsilon\dot{M}$ from $L_{\textrm{R}}$ and $M$\\
                           & $M_{\odot}\ \textrm{yr}^{-1}$           & $M_{\odot}\ \textrm{yr}^{-1}$                  \\
            \hline
            ACS ID \#12560 & $<\num{2.5e-17}$                       & $<\num{8.4e-15}$                              \\
            ACS ID \#21859 & $<\num{1.5e-17}$                       & $<\num{5.2e-15}$                              \\
            ACS ID \#5132  & $<\num{1.7e-17}$                       & $<\num{8.6e-15}$                              \\
            \hline
        \end{tabular}
    \end{table}

\subsubsection{Estimating the expected mass accretion rates through stellar wind capture}
    With the limits on $\epsilon\dot{M}$ shown in Table~\ref{tab:mdot_limits}, we can now place some constraints on the radiative efficiency of the accretion $\epsilon$ by making some assumptions as to what $\dot{M}$ is for these systems. A similar, but reversed calculation is shown in \citet{Breivik2019}, where they estimate $L_{\textrm{X}}$ based on assumptions over the wind mass-loss rate. For our case, as these systems are detached binary systems we expect that there will be no Roche-lobe overflow, as evidenced by the lack of an X-ray detection for any system. The formation of an accretion disc in a binary is dependent on the circularisation radius around the compact object. If the circularisation radius is smaller than the event horizon of the BH, no disc will be formed. In a stellar wind accretion regime, the circularisation radius can be expressed as \citep{Frank2002}
    \begin{equation}
        \frac{R_{\textrm{circ}}}{a} = \frac{M^3\qty(M+M_2)}{16\lambda^4(a)M^4_2} \qty(\frac{R_2}{a})^4,
    \end{equation}
    where $M$ is the compact object mass, $M_2$ and $R_2$ are the mass and radius of the companion respectively, and $a$ is the orbital separation of the binary. The term $\lambda(a)$ is the wind law describing the behaviour of the stellar winds from the companion. Due to the uncertainty over the wind law, we cannot confidently calculate the circularisation radii of the three BHs, as changing $\lambda(a)$ even by a factor of 2 drastically changes the result. Thus, we assume that the accretion present in these systems, if any, is due to the capture of the stellar winds from the companion stars with no accretion disc formed. We do note, however, that the BH ACS ID \# 21859 is the most likely system to form an accretion disc due to a shorter binary separation when compared to the other two BH systems.
    
    The amount of the wind captured by the compact object can be expressed as a fraction of the wind mass loss rate of the companion star \citep{Frank2002}
    \begin{equation}
        \frac{\dot{M}}{-\dot{M}_w} \cong \frac14 \qty(\frac{M}{M_2})^2\qty(\frac{R_2}{a})^2,
    \end{equation}
    where $-\dot{M}_w$ is the mass loss rate of the companion, $M$ is the compact object mass, $M_2$ and $R_2$ are the mass and radius of the companion respectively, and $a$ is the orbital separation of the binary. Substituting in the orbital separation $a$ from Kepler's Third Law reduces the above equation to
    \begin{align}
        \frac{\dot{M}}{-\dot{M}_w} &= \qty(\frac{\pi^4}{4G^2})^{1/3} \qty(\frac{R_2}{M_2})^2 \qty(\frac{M}{P})^{4/3}\\
        &\sim 0.01412 \qty(\frac{R}{R_{\odot}})^2 \qty(\frac{M_2}{M_{\odot}})^{-2} \qty(\frac{M}{M_{\odot}})^{4/3} \qty(\frac{P}{\si{days}})^{-4/3},
    \end{align}
    where $P$ is the binary orbital period. The mass accretion rate is dependent on the mass loss rate due to stellar winds from the companion stars.
    
    We adopt the relations from \citet{Johnstone2015,Johnstone2015a} to estimate the mass loss rate from the companion stars due to the stellar winds. The wind mass loss rate is expressed as
    \begin{equation}
        -\dot{M}_w = \dot{M}_{\odot} \qty(\frac{R}{R_{\odot}})^2 \qty(\frac{\Omega}{\Omega_{\odot}})^{1.33} \qty(\frac{M}{M_{\odot}})^{-3.36},
    \end{equation}
    where the solar wind mass loss rate $\dot{M}_{\odot}=\num{1.4e-14}M_{\odot}\ \textrm{yr}^{-1}$, the Carrington rotation rate $\Omega_{\odot}=\SI{2.67e-6}{\radian\per\second}$, and $R$, $\Omega$, and $M$ are the radius, rotational velocity, and mass of the companion star respectively. By knowing the radius, rotational velocity, and mass of the companion stars, we can estimate their wind mass loss rates. 
    
    The masses of the companion stars are given in \citet{Giesers2019}. The radii of the companions are derived based on the effective temperature and an estimate of the luminosity of the companion. The radii of the companions are: $R=1.62R_{\odot}$ for ACS ID\# 12560, $R=0.84R_{\odot}$ for ACS ID\# 21859, and $R=0.64R_{\odot}$ for ACS ID\# 5132. Through full-spectrum fits, the rotational velocities of the companions can be estimated, however, this is limited to fast rotators only due to the spectral resolution of MUSE. The rotational velocities measured are: $\Omega\sin i\leq\SI{2.42e-5}{\radian\per\second}$ for ACS ID\# 12560, $\Omega\sin i\leq\SI{5.12e-5}{\radian\per\second}$ for ACS ID\# 21859, and $\Omega\sin i=\SI{9.38e-5}{\radian\per\second}$ for ACS ID\# 5132. In the case of the first two sources, we only have upper limits on the rotational velocity of the companions as the true rotation rates are likely below the detection threshold of MUSE, and the constraint on ACS ID\# 5132 is only marginally above the detection threshold. Due to these uncertainties, we can use other theoretical models to predict what the actual rotation rates are.
    
    For the systems ACS ID\# 12560 and ACS ID\# 5132, we can use the gyrochronology predictions of \citet{Epstein2014} to provide a better constraint on the rotational velocity of the companions. Here, modern models predict that the rotational period will increase as $\textrm{age}^{-0.5}$, as predicted by \citet{Skumanich1972}, for stars older than 5 Gyr. For the companions in these two systems, we can expect rotation rates between 10 and 40 days, with an increase in period with stellar mass. Given the masses of the companions in ACS ID\# 12560 and ACS ID\# 5132, we adopt a predicted rotational period of $25\pm5$ days, implying a rotational velocitiy of $\Omega=\SI{2.9e-6}{\radian\per\second}$. However, it is important to note that the companion star in ACS ID\# 12560 is a sub-giant, and its rotational period will have been altered due to the expansion of its shell, likely resulting in a larger rotational period than an equally massive star on the main sequence. In this case, we note that applying the predictions of \citet{Epstein2014} may not give an appropriate estimate of the rotational period, and other gyrochronology models of sub-giant stars may need to be considered.

    For the case of the BH system ACS ID \#21859, due to the short orbital period of the system \citep[$\sim 2$ days,][]{Giesers2019} we can constrain the rotational velocity based on the assumption that the system is tidally locked. The location of this source in a GC means it is likely that the orbit was not originally circular due to its formation through dynamical channels. Given that the orbit of this source is now circular ($e\sim0.07$), it is reasonable to assume that the BH and companion are tidally locked, as tidal locking should occur prior to the circularisation of the binary orbit. We adopt the following equation for the rotational broadening for a star that is tidally locked from \citet{Wade1988}
    \begin{equation}
        V_{\textrm{rot}}\sin i \approx 0.462K_2 q^{1/3}(1+q)^{1/3},
    \end{equation}
    where $K_2$ is the semi-amplitude of the secondary and $q=M_2/M_1$. We do note that the caveat to using this relation to calculate the rotational velocity is that it assumes that the secondary is Roche-lobe filling, something that we have not assumed for our other calculations. From this, we estimate the rotational velocity of this star to be $\Omega\sin i=\SI{1.09e-4}{\radian\per\second}$.
    
    In Table~\ref{tab:mdot_calc} we present the fraction of the stellar wind captured by the BHs, the wind mass loss rates of the companions, and the final mass accretion rate into the BHs we estimate. As we expect, the mass accretion rates are low, consistent with there being no evidence of meaningful accretion in these systems. Notably, the mass accretion rate for ACS ID \#21859 is approximately five orders of magnitude higher than that of the other BH systems. This is due to this system having a much shorter binary orbital period by a factor of $\sim100$, meaning the BH is expected to capture a higher fraction of the stellar wind from the companion.
    
    \begin{table}
        \centering
        \caption{The mass accretion rate as a fraction of the stellar wind loss rate, the wind mass loss rate, and the mass accretion rate we estimate for each BH system. The mass accretion rate for ACS ID \#21859 is at least five orders of magnitude higher than that of the other BH systems due to its much shorter binary orbital period.}
        \label{tab:mdot_calc}
        \begin{tabular}{lccc}
            \hline
            \hline
            Source         & $\dot{M}/-\dot{M}_w$ & $-\dot{M}_w$                  & $\dot{M}$                     \\
                           &                      & $M_{\odot}\ \textrm{yr}^{-1}$ & $M_{\odot}\ \textrm{yr}^{-1}$ \\
            \hline
            ACS ID \#12560 & $\num{4.61e-4}$     & $\num{8.34e-14}$            & $\num{3.85e-17}$            \\
            ACS ID \#21859 & $\num{1.39e-1}$     & $\num{7.29e-12}$            & $\num{1.01e-12}$            \\
            ACS ID \#5132  & $\num{1.46e-5}$     & $\num{2.88e-14}$            & $\num{4.20e-19}$             \\
            \hline
        \end{tabular}
    \end{table}    

\subsubsection{Accretion efficiency constraints}    
    These accretion limits now allow us to place some constraints on what the radiative efficiency $\epsilon$ must be in these systems. Previous studies \citep[e.g., ][]{Maccarone2005} have assumed that the accretion flow is radiatively inefficient. This is the case with advection-dominated accretion flows \citep{Narayan1995}. For low accretion rates \citep[$\dot{M}/\dot{M_{\textrm{Edd}}}<0.02$,][]{Maccarone2003,VahdatMotlagh2019}, the efficiency scales with the accretion rate. This allows us to express the efficiency as
    \begin{equation} \label{eq:epsilon}
        \epsilon = 0.1\qty(\frac{\dot{M}}{\dot{M_{\textrm{Edd}}}}/0.02),
    \end{equation}
    where $\dot{M}_{\textrm{Edd}}$ is the Eddington accretion rate, and $\dot{M}$ is usually expressed as some fraction ($\sim0.03$) of the Bondi accretion rate based on the observations from \citet{Pellegrini2005}. 
    
    For the case of the detached BHs in NGC~3201, we assume that the mass accretion rate is the fraction of the stellar wind captured by the BH (the $\dot{M}$ values in Table~\ref{tab:mdot_calc}) with no accretion disc formed. Combining this with the accretion limits shown in Table~\ref{tab:mdot_limits}, we can constrain the radiative efficiency by assuming $\epsilon\dot{M}=limit$. These constraints on the radiative efficiency are shown in Table~\ref{tab:epsilon_limits}. The BH ACS ID \#21859 provides the best constraints on the radiative efficiency due to having a higher accretion rate from stellar winds. The lack of a radio and X-ray detection of this source indicates that it must be accreting below our sensitivity limits. The radio limit of this source constrains the efficiency to $<\num{5.1e-3}$, and the X-ray limit of this source constrains the efficiency to $<\num{1.5e-5}$. This provides a strong indication that either this source is not accreting, or is accreting extremely inefficiently for there to be no multiwavelength detection. This efficiency limit is consistent with what is theoretically expected from Equation~\ref{eq:epsilon} for low mass accretion rates.
    
    The constraints on the other two BHs are weaker, due to their lower expected accretion rates. Again, there is no multiwavelength detection of these sources, so the accretion must be below our sensitivity limits. Our radio observations are not deep enough to probe the expected accretion rates onto these BHs, and our X-ray observations only provide a constraint on the efficiency for ACS ID \#12560 of $<\num{6.5e-1}$. For ACS ID \#5132, the expected accretion rate is well below our sensitivity limits, so the radiative efficiency cannot be constrained.
    
    \begin{table}
        \centering
        \caption{The constraints on the radiative efficiency of the accretion onto the BHs assuming the accretion is in the form of stellar winds from the companion stars. The only system for which we have good constraints is ACS ID \#12560. The radiative efficiency of this system is less than $1\%$, indicating that the system is either not accreting, or the system is accreting extremely inefficiently. Entries consisting of a ``-'' indicate that no meaningful constraints on the radiative efficiency could be calculated due to the available depth of the radio and X-ray imaging.}
        \label{tab:epsilon_limits}
        \begin{tabular}{lcc}
            \hline
            \hline
            Source         & $\epsilon$ from $L_{\textrm{X}}$ & $\epsilon$ from $L_{\textrm{R}}$ and $M$\\
            \hline
            ACS ID \#12560 & $<\num{6.5e-1}$                         & -\\
            ACS ID \#21859 & $<\num{1.5e-5}$                     & $<\num{5.1e-3}$ \\
            ACS ID \#5132  & - & - \\
            \hline
        \end{tabular}
    \end{table} 
    
    Of course, these limits on the radiative efficiency are heavily dependent on the assumptions we have made, in particular that the accretion is purely from capturing a fraction of the companion's stellar winds with no accretion disc being formed, and the models we have assumed to calculate the wind mass loss rate from the stars. Furthermore, the rotational velocity of the source ACS ID \#21859 has been calculated assuming the star is tidally locked with the binary orbit, and the rotational velocities of the other two systems have been predicted through gyrochronology. Any change in these assumptions would alter the constraints placed on the radiative efficiency. However, this is the first time that constraints have been placed on the accretion efficiency for dynamically-confirmed stellar-mass BHs in a GC, and these show that any accretion is extremely inefficient. Any multiwavelength emission from this accretion would be very faint and would require large integration times to detect.

\subsection{Detectability of the black holes with current and future instruments}
    The upper limits on the radio and X-ray luminosities of the BHs in NGC~3201 also allow us to comment on the detectability of these systems with current instruments. Plotting these limits on the radio--X-ray luminosity plane (Figure~\ref{fig:lrlx}) shows that these sources lie well above the standard track occupied by accreting BHs \citep[$L_{\textrm{R}}\propto0.6L_{\textrm{X}, }$][]{Gallo2014}. This indicates that while our X-ray observations of the cluster may be deep enough to potentially probe faint emission from weakly accreting systems, our current radio observations are too shallow by at least one order of magnitude. 
    
    We can now comment on whether these limits could potentially be reachable through observations with current radio facilities, such as the ATCA. The BH ACS ID \#21859 has the deepest 1-10 keV X-ray limit of \SI{8.6e29}{\erg\per\second}. If we assume that any X-ray emission is on the verge of detectability with current data, the source falls on the standard accreting BH correlation, and that this correlation holds for these very low luminosities, the corresponding radio limit for this source is \SI{7.7e25}{\erg\per\second}, implying a 5.5 GHz radio flux density of \SI{5.2e-1}{\micro\jansky} assuming a cluster distance of 4.74 kpc. The current $3\mhyphen\sigma$ radio upper limit for this BH is \SI{9.5}{\micro\jansky} from 18.1 hr of ATCA observations, a value approximately 17 times too shallow to reach even the radio limit of the standard accreting BH correlation. As RMS noise in radio images decreases as $\sqrt{\textrm{time}}$, one would need to observe NGC~3201 for over 5200 hr of ATCA, observations assuming ideal conditions, to potentially detect emission at this limit from the system. This is not feasible for current generation instruments. Sub-\si{\micro\jansky} radio noise levels would be possible for some clusters with a few hundred hours of ATCA observations assuming the dynamic range of the observations is not limited, however, current instruments are not suited to probe the faintest BHs at these very low luminosities. These deep surveys, with noise levels on the order of hundreds of \si{\nano\jansky}, will be possible with next generation radio facilities such as the Next Generation VLA and the Square Kilometre Array \citep{Murphy2018,Dewdney2009}.
    
    Additionally, observations of these faintly accreting sources may be feasible with the proposed next generation of X-ray facilities, such as the \textit{Athena} X-ray Observatory \citep{Nandra2013}, and particularly the \textit{Lynx} X-ray Observatory \citep{TheLynxTeam2018} due to its angular resolution. To push our current X-ray limits deeper with our current X-ray facilities, $>100$ ks of \chandra observations would be needed.

\subsection{Population of X-ray sources and encounter rate} \label{sec:xrbs}
    Given we now have an estimate as to how many XRBs we expect in NGC~3201, we can now compare it to the stellar encounter rate \citep{Verbunt1987} of the cluster ($\Gamma\propto\int\rho^2/\sigma$) to see if there is a significant (over)under-population of XRBs. The population of XRBs is expected to be different depending on whether the cluster has undergone a core-collapse, and whether the cluster has a larger or smaller stellar encounter rate (i.e. how dynamically active is the cluster). It has previously been shown that more XRBs are seen in clusters with higher stellar encounter rates \citep{Heinke2003a,Pooley2003,Bahramian2013}. It appears that core-collapsed clusters have fewer XRBs than non-core-collapsed clusters for the same encounter rate \citep{Fregeau2008,Bahramian2013}.

    \begin{figure}
        \centering
        \includegraphics[width=\columnwidth]{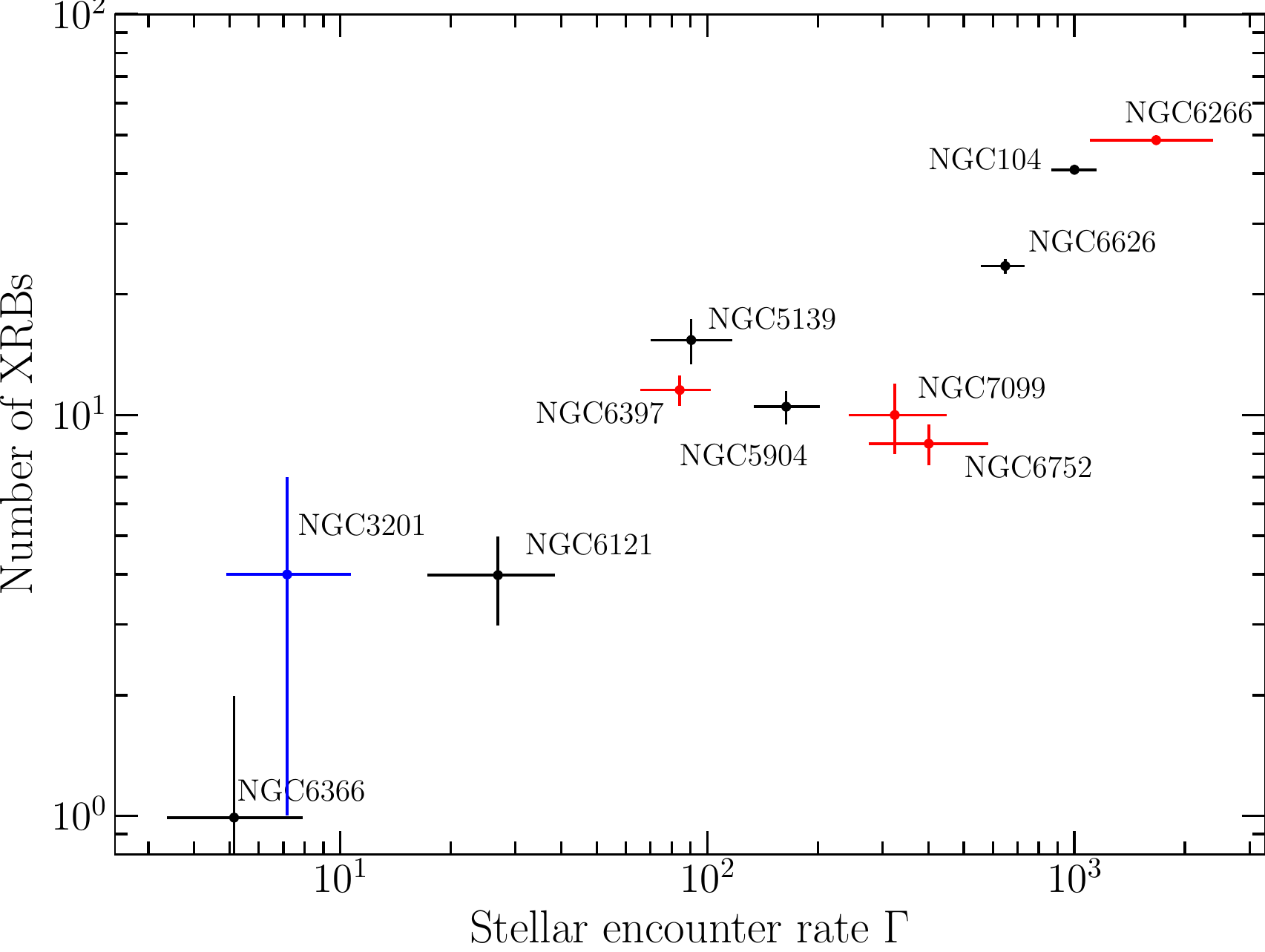}
        \caption{The number of XRBs plotted against the stellar encounter rate for a number of GCs. The clusters shown in red are core-collapsed clusters. NGC~3201 is shown in blue. The stellar encounter rate has been normalised so that the encounter rate of 47 Tucanae (NGC~104) is 1000.}
        \label{fig:xrbvsgamma}
    \end{figure}
    
    Figure~\ref{fig:xrbvsgamma} shows a plot of the number of XRBs in a cluster plotted against the stellar encounter rate of the cluster for a number of GCs. The number of XRBs for each cluster was collated from \citet{Pooley2003}, and only includes X-ray sources that are brighter than \SI{4e30}{\erg\per\second}. 
    
    For NGC~3201, we are confident that one X-ray source is an XRB an associated with the cluster. This source is the candidate RS, CXOU J101730.49-462442.4. Beyond this, three of the optical/X-ray sources discussed in \S~\ref{sec:optical_x-rays} are potential XRBs. Using the relations from \citet{Georgakakis2008}, we expect there to be approximately 14 extra-galactic X-ray sources within the half-light radius of the cluster with a luminosity greater than \SI{4e30}{\erg\per\second}. To estimate the most likely number of XRBs in the cluster, we assumed the number of background AGNs detected given the flux threshold follow a Poisson distribution with a Poisson rate corresponding to the 14 background sources expected within the cluster half-light radius, and that the likely number of XRBs cannot be less than the number of XRBs confirmed via multiwavelength observations. Thus, we estimate the likely number of XRBs to be $4\pm3$, based on the mode of the resulting distribution and 68\% interval, as indicated in Figure~\ref{fig:xrbvsgamma}.
    
    In Figure~\ref{fig:xrbvsgamma}, we see that NGC~3201 potentially contains more XRBs than would be expected for a cluster with a similar encounter rate. NGC~6366, which has an encounter rate of $\Gamma\sim5.1$ compared to $\Gamma\sim7.2$ for NGC~3201, only contains one XRB while NGC~3201 potentially contains four or more XRBs. This is unusual as for these low encounter rate clusters we do not expect many XRBs to form, as one of the main formation channels, dynamical interactions, is mitigated due to the low encounter rate. The potential overabundance suggests we could be observing primordial binary systems, formed towards the beginning of the cluster's life cycle, instead of being formed dynamically throughout the lifetime of the cluster. It is not unfeasible for primordial binaries to survive in a cluster such as NGC~3201. The low encounter rate of the cluster combined with its large mass \citep[$\num{1.6e5} M_{\odot}$;][]{Baumgardt2019} when compared to NGC~6366 \citep[$\num{3.8e4} M_{\odot}$;][]{Baumgardt2019}, means that it is less likely for these systems to be disrupted, as opposed to a more dynamically active cluster such as 47 Tucanae (NGC 104), where the higher encounter rate would make it difficult for primordial binaries to survive for long periods of time.
    
    However, given that the lower limit for the number of XRBs is consistent with that of NGC~6366, it is also possible that NGC~3201 displays no overabundance of XRBs at all. If this is the case, then NGC~3201 would be displaying a number of XRBs which is expected for its stellar encounter rate. This would indicate that from an XRB perspective, NGC~3201 is a standard GC.


\subsection{Evolution of sub-subgiants} \label{sec:ssg_disc}
    In NGC~3201 we detect two of the SSGs identified in the MUSE binary catalogue as X-ray sources in the MAVERIC survey, and provide X-ray upper limits for the two other SSGs. We also identify an optical source with X-ray emission, which we conclude to be a RS system based on its position in the cluster CMD. This means that there is now a total of five systems in NGC~3201 that lie redward of the main sequence and are underluminous compared to normal (sub-)giant stars. 
    
    The X-ray luminosities of these systems are consistent with the expected luminosities of SSG systems. SSGs are expected to have very low X-ray luminosities on the order of $10^{30-31}$ \si{\erg\per\second} \citep{Geller2017}, and the three sources with confident X-ray detections have X-ray luminosities within this range. Beyond the X-ray luminosities, the orbital periods of three of the four SSGs detected by MUSE are also consistent with the expected orbital periods for SSG systems, which have been observed to be $\lesssim15$ days \citep{Geller2017}. The SSG with a longer orbital period, ACS ID \#11405 with a period of $\sim17$ days, also has the largest eccentricity of any SSG, potentially indicating that it is in an earlier stage of its evolution prior to the orbit circularising, either through mass transfer or tidal forces.
    
    Finally, the observed properties of these systems are also somewhat consistent with some speculative formation pathways for SSGs. \citet{Leiner2017} propose three methods for SSG formation: binary mass transfer, stripping of the star's envelope during a potential dynamical encounter, and strong magnetic fields causing underluminous stars. Binary mass transfer in the form of accretion onto a compact object would produce an X-ray signal, which we observe from two SSGs and our candidate RS indicating some accretion may be present in these systems. Another feasible explanation to the X-ray emission seen is that these two SSGs may be ABs, given that their X-ray luminosities fall in the range expected for these systems \citep[$\lesssim10^{31}$ \si{\erg\per\second};][]{Gudel2002}. All five of these systems are also underluminous in the optical bands when compared to normal (sub-)giant stars. It has also been suggested that the presence of underluminous stars in GCs could be indicative of BH presence \citep{Ivanova2017}, a statement that now has more merit given the presence of BHs in NGC~3201.

\section{Conclusions}
    In this paper, we present a catalogue of energetic sources in NGC~3201. We combine the radio and X-ray data from the MAVERIC survey \citep{Shishkovsky2020,Bahramian2020} with the spectral binary and emission line catalogues of the cluster produced by MUSE \citep{Giesers2019,Gottgens2019} and the HUGS catalogue \citep{Piotto2015,Nardiello2018} to investigate any binary system or optical source that has radio and/or X-ray emission. From this, we consider 42 sources in this paper. Three sources are known (or candidate) BHs and four sources are known SSGs \citep{Giesers2019}. Within the cluster half-light radius we also identify a new candidate red straggler system, four optical sources with X-ray emission, two sources with radio and X-ray emission, 15 X-ray sources with no other multiwavelength counterpart, and 13 radio sources with no other multiwavelength counterpart. 
    We speculate that three of the four optical sources with X-ray emission are some type of X-ray emitting binary (e.g. AB, CV, or XRB) and, along with the candidate red straggler, are associated with the cluster. We suspect that the remaining radio and X-ray sources can be explained as background sources.
    
    Importantly, we present the first radio and X-ray limits on the detached BHs in NGC~3201. All three sources are not detected in the 5.5 GHz radio band or the 0.5-10 keV X-ray band. The lack of a multiwavelength counterpart to these systems suggests that these systems are not accreting. From these limits, we are able to provide some constraints on the radiative efficiency of any accretion that is present in these, assuming that any accretion is through the capture of the companions' stellar winds. These calculations suggest that the accretion in these systems is extremely inefficient, with the limits on the BH ACS ID \#21859 suggesting that the radiative efficiency is $\lesssim\num{1.5e-5}$, consistent with previous work suggesting that efficiency scales with accretion rate for low accretion rates \citep[e.g., ][]{Maccarone2003,VahdatMotlagh2019}. Due to the extremely faint nature of these systems ($L_{\textrm{X}}<10^{30}\si{\erg\per\second}$), it is quite challenging to detect them with the current generation of radio facilities. Assuming that these BH systems follow the standard track for accreting BHs \citep{Gallo2014}, hundreds of observing hours would be required by either the ATCA or the VLA to reach the sub-\si{\micro\jansky} noise levels needed to probe any potential radio emission from these sources. However, these weakly accreting targets are prime candidates for deep surveys with the next generation of radio facilities, such as the Next Generation VLA or the Square Kilometre Array, and observations of these sources may be feasible with the next generation of X-ray facilities.
    
    NGC~3201 may also have a slight overpopulation of XRBs when compared to other GCs with similar stellar encounter rates. We speculate that this could be due to the presence of primordial binaries that have not been disrupted. However, if this overabundance is not true, then NGC~3201 contains a number of XRBs that would be expected for its stellar encounter rate.

\section*{Acknowledgements}
    We thank the anonymous referee for their helpful comments on this manuscript. AP was supported by an Australian Government Research Training Program (RTP) Stipend and RTP Fee-Offset Scholarship through Federation University Australia. JS acknowledges support from NASA grant 80NSSC21K0628 and the Packard Foundation. SK gratefully acknowledges funding from UKRI in the form of a Future Leaders Fellowship (grant no. MR/T022868/1). COH is supported by NSERC Discovery Grant RGPIN-2016-04602. The Australia Telescope Compact Array is part of the Australia Telescope National Facility which is funded by the Australian Government for operation as a National Facility managed by CSIRO. We acknowledge the Gomeroi people as the traditional owners of the Observatory site. The National Radio Astronomy Observatory is a facility of the National Science Foundation operated under cooperative agreement by Associated Universities, Inc.. We acknowledge the use of the following packages/software in this work: \textsc{ciao} \citep{Fruscione2006}, provided by the \chandra X-ray centre (CXC); \textsc{heasoft}, provided by the High Energy Astrophysics Science Archive Research Centre (HEASARC); \textsc{topcat} \citep{Taylor2005}; and the \textsc{python} packages \textsc{astropy} \citep{AstropyCollaboration2018}, \textsc{matplotlib} \citep{Hunter2007}, \textsc{numpy} \citep{Harris2020}, and \textsc{the joker} \citep{Price-Whelan2017}. This work made use of NASA's Astrophysics Data System and arXiv.

\section*{Data Availability}
    The data underlying this work are available in the following locations: X-ray data are available from the \chandra Data Archive at \url{https://cda.harvard.edu/chaser/}, and the reduction of this data is outlined in \citet{Bahramian2020}; radio data are available from the Australia Telescope Online Archive (ATOA) at \url{https://atoa.atnf.csiro.au/}; optical spectroscopic data from MUSE are available from the ESO Science Archive Facility at \url{http://archive.eso.org/}, and the reduction and results from this data are outlined in \citet{Giesers2018,Giesers2019}; and optical photometric data from the HUGS catalogue are available at \url{https://archive.stsci.edu/prepds/hugs/}, with the details of this catalogue detailed in \citet{Piotto2015,Nardiello2018}.



\bibliographystyle{mnras}
\bibliography{all_references} 



\bsp	
\label{lastpage}
\end{document}